\documentclass[fleqn,10pt]{wlscirep}

\usepackage[utf8]{inputenc} %
\usepackage[T1]{fontenc}    %
\usepackage{hyperref}       %
\usepackage{url}            %
\usepackage{booktabs}       %
\usepackage{amsfonts}       %
\usepackage{nicefrac}       %
\usepackage{microtype}      %
\usepackage{xcolor}         %
\usepackage{amsmath}
\usepackage{siunitx}
\usepackage{graphicx}
\usepackage{makecell}
\newcommand*{\defeq}{\mathrel{\vcenter{\baselineskip0.5ex \lineskiplimit0pt
                     \hbox{\scriptsize.}\hbox{\scriptsize.}}}%
                     =}
\usepackage{bm}
\usepackage[
    natbib=true,
    style=numeric-comp,
    sorting=none
    ]{biblatex}
\usepackage{booktabs}
\usepackage{multirow}
\usepackage{bbm}

\usepackage{algorithmic}
\usepackage{algorithm}
\usepackage{rotating}

\DeclareUnicodeCharacter{0301}{\'{e}}

\title{State-specific protein-ligand complex structure prediction with a multi-scale deep generative model}

\author[1,*]{Zhuoran Qiao}
\author[3]{Weili Nie}
\author[3]{Arash Vahdat}
\author[1,4]{Thomas F. Miller III}
\author[2,3]{Animashree Anandkumar}
\affil[1]{Division of Chemistry and Chemical Engineering, California Institute of Technology, Pasadena, CA 91125}
\affil[2]{Division of Engineering and Applied Science, California Institute of Technology, Pasadena, CA 91125}
\affil[3]{Nvidia Corporation, Santa Clara, CA 95051}
\affil[4]{Entos, Inc., Los Angeles, CA 90027}

\affil[*]{Current e-mail address: zhuoran.qiao@entos.ai. }

\begin{document}

\begin{abstract}
The binding complexes formed by proteins and small molecule ligands are ubiquitous and critical to life. %
Despite recent advancements in protein structure prediction, existing algorithms are so far unable to systematically predict the binding ligand structures along with their regulatory effects on protein folding.
To address this discrepancy, we present NeuralPLexer, a computational approach that can directly predict protein-ligand complex structures solely using protein sequence and ligand molecular graph inputs. 
NeuralPLexer adopts a deep generative model to sample the three-dimensional (3D) structures of the binding complex and their conformational changes at an atomistic resolution. %
The generative model is based on a diffusion process that incorporates essential biophysical constraints and a multi-scale geometric deep learning system to iteratively sample residue-level contact maps and all heavy-atom coordinates in a hierarchical manner. 
NeuralPLexer achieves state-of-the-art performance compared to all existing %
methods on benchmarks for both protein-ligand blind docking and flexible binding site structure recovery. Moreover, owing to its specificity in sampling both ligand-free-state and ligand-bound-state ensembles, NeuralPLexer consistently outperforms AlphaFold2 in terms of global protein structure prediction accuracy
on %
both representative structure pairs with large conformational changes (average TM-score=0.93) and recently determined ligand-binding proteins (average TM-score=0.89). 
Case studies reveal that the predicted conformational variations are consistent with structure determination experiments for important targets, including human KRAS\textsuperscript{G12C}, ketol-acid reductoisomerase, and purine GPCRs.
Our study suggests that a data-driven approach can capture the structural cooperativity between proteins and small molecules, showing 
promise in accelerating the design of enzymes, drug molecules, and beyond. 
\end{abstract}

\flushbottom
\maketitle

\thispagestyle{empty}

\begin{figure}
    \centering
    \includegraphics[width=\textwidth]{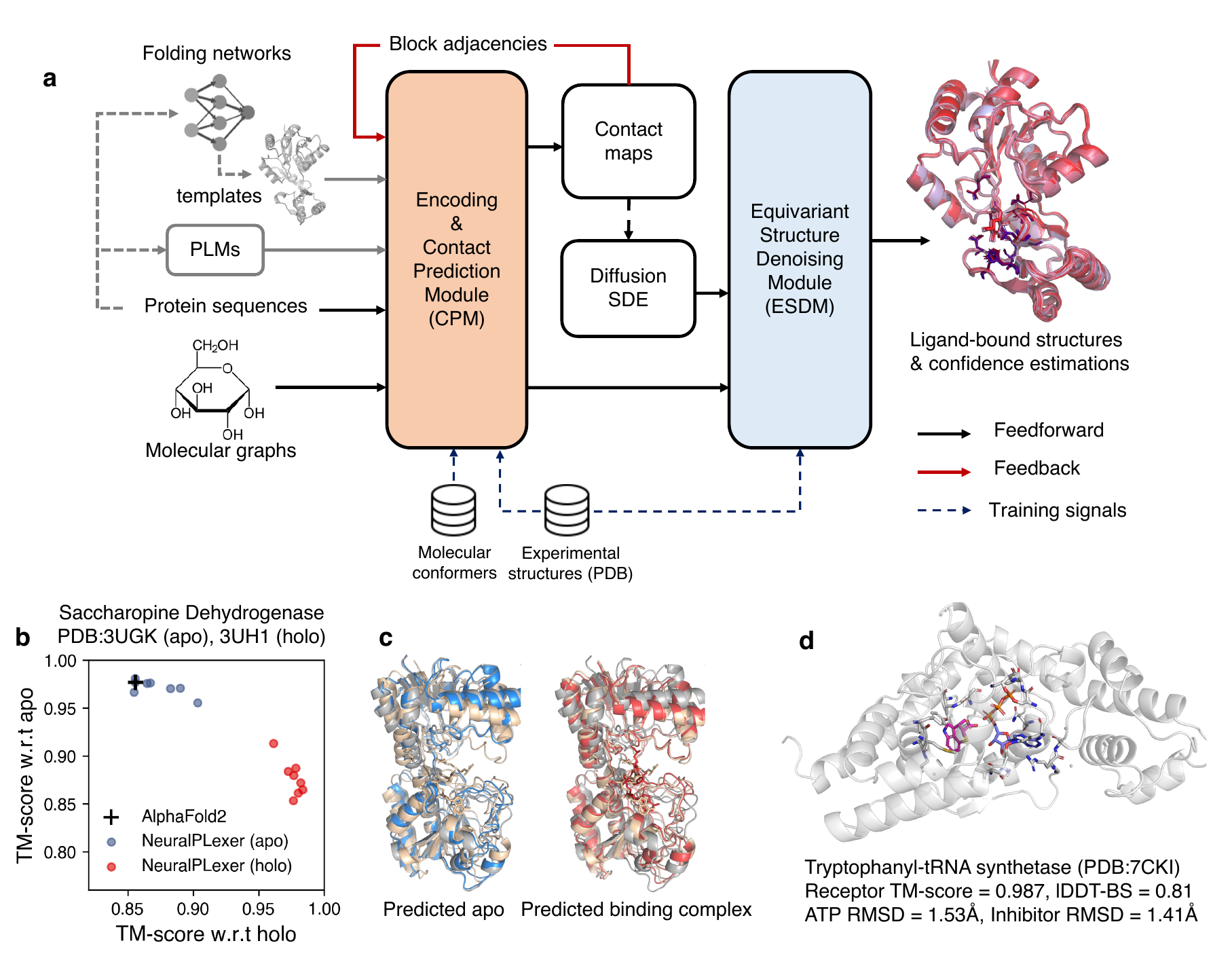}
    \caption{NeuralPLexer enables accurate prediction of protein-ligand complex structure and conformational changes. (a) Method overview. To perform predictions, the input protein sequence is first used to retrieve protein language model (PLM) features and structure templates; NeuralPLexer then combines the set of PLM and template features with molecular graph representations of the input ligands to directly sample an ensemble of binding complex structures via a multi-scale generative model. The main network of NeuralPLexer is comprised of a coarse-grained, auto-regressive contact prediction module (CPM) and an atomistic, diffusion-based equivariant structure denoising module (ESDM). SDE: stochastic differential equation. (b-c) Prediction example on a target with large-scale domain motions upon ligand binding (UniProt:P38998). (b) The structure similarities against experimental apo (i.e., ligand-free protein, PDB:3UGK) and holo (i.e., ligand-bound protein, PDB:3UH1) structures measured by TM-score are plotted for AlphaFold2 predictions (grey), ligand-free NeuralPLexer predictions (blue), and ligand-bound NeuralPLexer predictions (red). (c) Visualizations of representative NeuralPLexer-predicted structures (blue for apo, red for holo) are overlaid with the experimental apo structure (grey) and the holo structure (light yellow). (d) Visualization of a prediction example (PDB:7CKI, UniProt:P00953) for which NeuralPLexer achieves high atomic accuracy for both the ATP (blue) and an inhibitor bound to the tryptophan site (magenta) upon an induced-fit structure rearrangement.
    }
    \label{fig:fig1}
\end{figure}

\section*{Main}

Deep learning has enabled great advancements in predicting protein structures from their one-dimensional amino acid sequences. State-of-the-art protein structure prediction networks such as AlphaFold2~\cite{jumper2021highly} (AF2) adopt prediction pipelines based on evolutionary, physical, and geometric constraints of protein structures. Specifically, evolutionary constraints extracted from multiple sequence alignments (MSAs)~\cite{ovchinnikov2017protein,yang2020improved,baek2021accurate,baek2022accurate} or protein language models (PLMs)~\cite{chowdhury2022single,wang2022single,wu2022high,lin2023evolutionary} and specialized neural networks are systematically combined with the sequence-based information and geometrical representations to enable end-to-end 3D structure prediction.
Despite remarkable success in predicting protein crystal structures, such a single-structure formulation of the protein folding problem provides incomplete information about protein function. It is also often found insufficient for structure-based drug design~\cite{zhang2022benchmarking,wong2022benchmarking}.
As discussed by numerous studies~\cite{henzler2007dynamic,popovych2006dynamically,nussinov2013allostery,ayaz2023,lane2023protein}, protein structures are dynamically modulated by their interactions with small molecules such as ligands and post-translational modifications. 
Such dynamical protein conformational changes occur on vastly varied kinetic timescales ranging from nanosecond-scale vibrations to millisecond-scale collective motions, and can trigger diverse downstream responses that are crucial to the regulation of biological functions. 
Proposing ligands that selectively target protein conformations has also become an increasingly important strategy in small molecule-based therapeutics~\cite{lawson2012antibody,moore2020ras,draper2021positive}. 

Nevertheless, the computational modeling of protein-ligand complexes that are coupled to significant changes in receptor conformation is hampered by the prohibitive cost of simulating slow protein state transitions~\cite{shaw2010atomic,shan2022does}. %
In the past decades, several schemes have been proposed to remedy these issues; those include methods based on molecular dynamics simulation and enhanced sampling techniques~\cite{sinko2013accounting,zhao2021enhanced,zhang2022benchmarking,vani2022sequence}, molecular docking guided by template-based modeling~\cite{abagyan1994icm,wu2018coach,hekkelman2023alphafill} and iterative refinement protocols~\cite{fischer2014incorporation,ollikainen2015coupling,amaral2017protein,wang2022scaffolding}, as well as recently proposed modifications to structure prediction networks~\cite{stein2021modeling,del2022sampling,heo2022multi,sala2022biasing}. %
However, such methods often require case-specific expert interventions or constraints from experimental data, and are still not a unified framework to systematically predict binding complex structures at a proteome scale.
Moreover, the applicability of existing deep-learning-based algorithms to ligand-binding proteins is limited by their single structure regression-based formulations, especially for non-endogenous small molecule binding~\cite{goodman1996goodman} for which the ligand identity cannot be inferred from protein motifs. 

Recent developments in generative deep learning have provided an alternative paradigm, and they have demonstrated substantial progress in understanding complex vision and language domains~\cite{dhariwal2021diffusion,brown2020language,ramesh2022hierarchical,saharia2022photorealistic}. 
Two notable strategies for generative modeling include (a) auto-regressive models as widely adopted in transformer~\cite{vaswani2017attention} networks for sequence data such as natural language and genomic~\cite{Zvyagin2022.10.10.511571,avsec2021effective} 
sequences based on a sequential sampling process from learned token distributions; and (b) diffusion-based generative models that leverage a stochastic process to generate data by sampling from a prior distribution and using a neural network to progressively reverse the noising process. 
Several works have demonstrated that deep generative models are capable of producing de novo-designed proteins with experimentally validated functions, for example, some works use language models for protein sequence design (such as ESM~\cite{bepler2021learning,lin2023evolutionary} and ProteinMPNN~\cite{dauparas2022robust}) and others have applied diffusion models for protein backbone generation~\cite{Ingraham2022.12.01.518682,rfdiffusion,wu2022protein,lin2023generating}. Despite so, these works have not accounted for small molecules when generating protein sequences or 3D structures. 
Other studies~\cite{xu2022geodiff,jing2022torsional,tankbind,nakata2022end,corso2022diffdock,DPL,diffsbdd} have shown that diffusion models are effective in modeling molecular structures beyond protein backbones, especially in the context of molecular docking and structure-based drug design.  
However, no work has so far developed generative models capable of directly predicting binding complex structures at an atomistic resolution and with an accuracy comparable to structure determination experiments. 

In this study, we conclusively tackle this challenge through a deep generative model informed by biophysical inductive biases. We demonstrate that the integration of these inductive biases into a deep neural network, which combines both auto-regressive and diffusion generative modeling, is key to accurately predicting protein-ligand complex structures at scale. This is aligned with two essential factors that dictate ligand binding: (a) the determination of the global contextual information related to ligand function, such as selectivity to orthosteric or allosteric sites, and (b) the process of resolving energetically favorable inter-atomic structures based on sub-nanometer-scale physical interactions.

Our framework, known as NeuralPLexer, is a computational system that predicts protein-ligand complex structures using an end-to-end generative modeling strategy. 
The presented method directly generates a structural ensemble of binding complexes given protein sequence and ligand molecular graphs inputs, conditioning on auxiliary features obtained from PLMs and template protein structures retrieved from experimentally resolved homologs or computational models.
Both the prediction pipeline and the underlying neural network architecture are designed to mirror the multi-scale hierarchical organization of biomolecular complexes. 
In particular, NeuralPLexer comprises of (a) a graph-based network to encode the atomic scale chemical and geometrical features of individual small molecules and amino-acid graphs into tensor representations, enabled by a physics-inspired network architecture that is trained on million-scale molecular conformation and bio-activity databases; (b) a contact prediction module (CPM) to generate residue-scale inter-molecular distance distributions, coarse-grained contact maps, and associated pair representations using an attention-based network, motivated by recent vision-language models and folding prediction networks~\cite{alayrac2022flamingo,jumper2021highly}; and (c) an equivariant structure denoising module (ESDM) to generate the binding complex atomistic structure conditioned on outputs from the atomic-scale and residue-scale networks, using a structured denoising diffusion process that is equivariant and preserves chirality constraints of the protein and ligand molecules. 

When benchmarked on blind protein-ligand docking, NeuralPLexer improves the ligand pose accuracy by up to 78\% compared to the best-performing existing method on the PDBBind2020 dataset~\cite{wang2005pdbbind}. In the context of ligand binding site design, where to our knowledge no existing machine learning approaches are applicable, NeuralPLexer effectively recovers up to 46\% of binding site structures by solely using truncated scaffolds generated from AF2. This represents an at least 59\% improvement in success rate when compared to the physics-based method in Rosetta~\cite{davis2009rosettaligand}. 
On two carefully-curated benchmark datasets of ligand-binding proteins with large structural plasticity, NeuralPLexer outperforms the state-of-the-art protein structure prediction algorithm AF2 as demonstrated by the highest TM-score~\cite{zhang2005tm} (0.906 on average) and by 11-13\% improvements in terms of accuracy on domains that undergo substantial conformational changes upon ligand binding. 
The above results suggest that our strategy enjoys a systematic advantage over existing methods in selectively predicting protein structures that are subject to induced-fit binding or conformational selection. 
The versatile capability of NeuralPLexer to model ligand binding and protein structural variations can enable rapid characterization of conformational landscapes to facilitate a better understanding of molecular mechanisms governing protein functions, thus aiding the proteome-scale identification of unconventional targets for therapeutic interventions and protein engineering. 

\section*{Results}

\subsection*{NeuralPLexer}
\label{method}

\begin{figure}[htb]
    \centering
    \includegraphics[width=\textwidth]{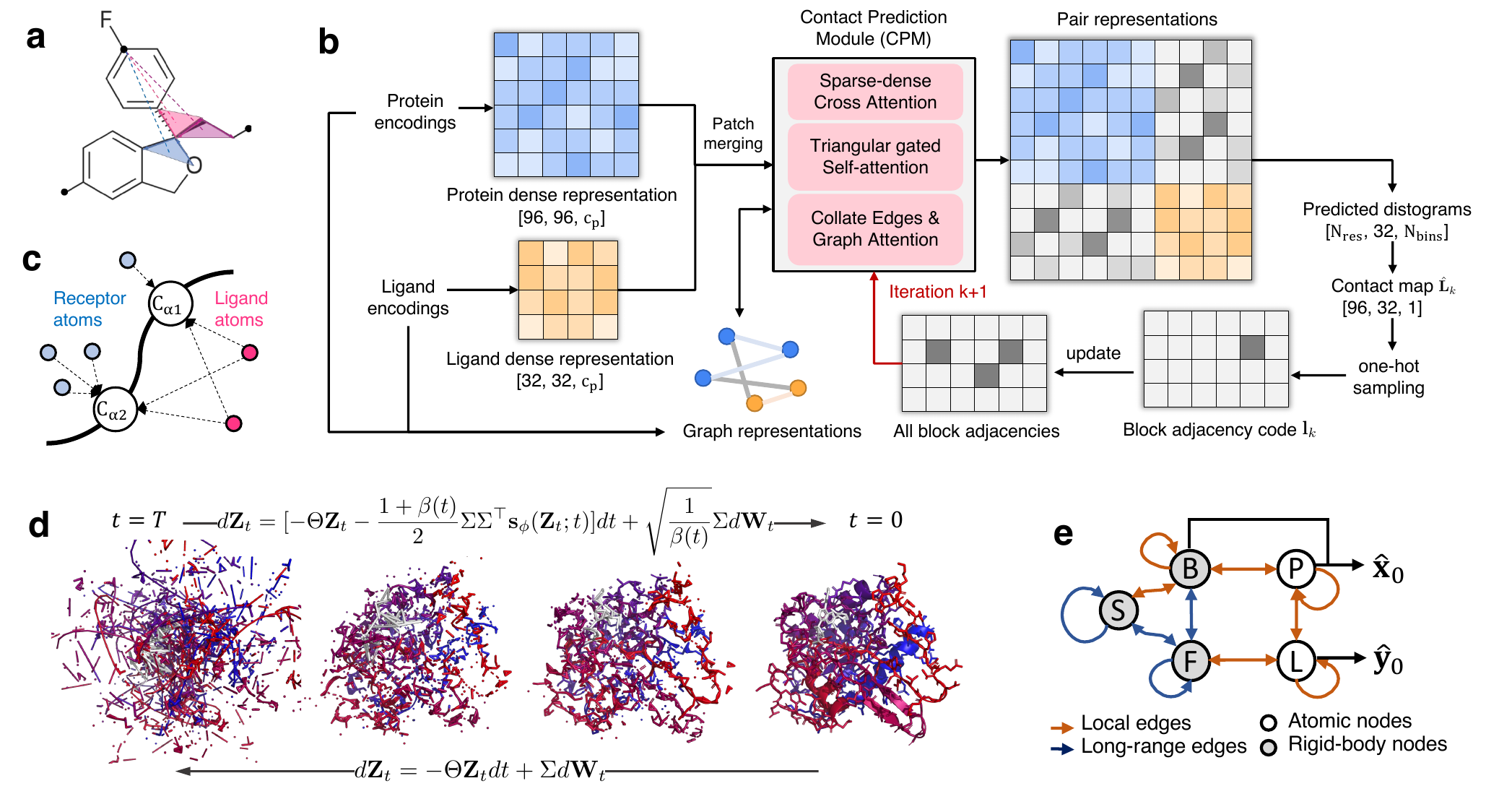}
    \caption{Architecture details. 
    (a) Ligand molecules and amino acids are encoded as the collection of atoms, local coordinate frames (depicted as semi-transparent triangles), and stereochemistry-specific pairwise embeddings (depicted as dashed lines) representing their interactions. (b) Information flow in the contact prediction module (CPM) network. The CPM network samples block-wise adjacency matrices among protein and ligand nodes using an autoregressive decoding scheme, where the block adjacency matrices $\mathbf{l}_k$ sampled from last step are passed to the network to update the predicted histograms of pairwise distances ("distograms") and contact maps $\hat{\mathbf{L}}$. (c) The forward-time SDE introduces structured drift and noise covariance terms among protein $\mathrm{C\alpha}$ atoms, non-$\mathrm{C\alpha}$ atoms, and ligand atoms. %
    (d) Denoising diffusion process to generate the binding complex 3D atomistic structure. The protein (colored as red-blue from N- to C-terminus) and ligand (colored as grey) structures are jointly generated through a reverse-time, simulated annealing SDE starting from randomly initialized coordinate variables. 
    (e) Information flow in the ESDM neural network. The ESDM network operates on a heterogeneous graph formed by protein atoms (P), ligand atoms (L), protein backbone frames for all residues (B), backbone frames of the selected patches (S), and ligand local frames (F) to predict the denoised atomic coordinates $\hat{\mathbf{x}}_0, \hat{\mathbf{y}}_0$ used in the reverse-time SDE. The heterogeneous graph comprises randomized local edges (orange arrows) and densely connected long-range edges (blue arrows), where the long-range edges and inter-residue local edges are initialized via the CPM embeddings. %
    }
    \label{fig:fig2}
\end{figure}

The NeuralPLexer network architecture and the information flows during model training and inference are illustrated in Figure~\ref{fig:fig1}a. %
The primary model inputs are a set of protein amino acid sequences $\{ \mathbf{s} \}$ (containing a single chain for monomers and multiple chains for protein complexes)
and, when the binding ligands (one or multiple molecules) are present, a set of molecular graphs $\{G\}$ containing atomic numbers, bond types and stereochemistry labels (e.g., tetrahedral or E/Z isomerism~\cite{eliel1994stereochemistry}). To perform structure prediction, NeuralPLexer jointly samples %
the 3D heavy-atom coordinates of the protein $\mathbf{x}$ and those of the ligands $\mathbf{y}$ from a generative model conditioned on the sequence and graph inputs $\{ \mathbf{s} \}, \{G\}$. 
In addition to the primary sequence and graph inputs, we retrieved inputs from readily available transformer protein language models and templates from alternative experimental structures or protein structure prediction networks to provide extra conditioning signals to the generative model. In particular, we use protein sequence embeddings from the ESM-2~\cite{lin2023evolutionary} language model and template structures generated from AlphaFold2 as auxiliary inputs in this study.

During inference, the input protein sequences, retrieved auxiliary protein embeddings, and molecule graphs are first encoded and passed into a contact prediction module (CPM) that auto-regressively generates the proximity distributions and associated pairwise embeddings. %
An equivariant structure denoising module (ESDM) then generates the 3D structures by denoising the atomic coordinates %
using a learned stochastic process (Figure~\ref{fig:fig2}d). %
Finally, for every sampled structure a predicted confidence score (pLDDT, Supplementary Information~\ref{sec:si_plddt}) is assigned to each protein residue and ligand atom. A more technical summary of the model inference procedure is described in Supplementary Information, Algorithm~\ref{alg:main}.

\paragraph{Molecular representations}
To enable stereospecific molecular geometry representation and explicit reasoning about long-range geometrical correlations, NeuralPLexer hybridizes two types of elementary molecular representations (Figure~\ref{fig:fig2}a): (a) atom nodes and (b) rigid-body nodes representing coordinate frames formed by adjacent chemical bonds (hereby "frames"). We introduce a pre-trained encoder model termed  Molecular Heat Transformer (MHT; Methods, molecular graph encoding) to transform the ligand and amino acid molecular graphs into a set of informative embeddings;
once the MHT outputs are gathered for all ligand molecular graphs, we uniformly select $N_\mathrm{L}$ frames and use the edge embeddings among the selected frames to create a dense tensor representation $\mathbf{F}_\mathrm{L}\in \mathbb{R}^{N_\mathrm{L}\times N_\mathrm{L} \times c_\mathrm{p}}$ of all input ligands, where $c_\mathrm{p}$ is the embedding dimension of the network. Parallel to this molecular graph branch, the input protein sequences and auxiliary features are encoded into a tensor representation $\mathbf{F}_\mathrm{P}\in \mathbb{R}^{N_\mathrm{P}\times N_\mathrm{P} \times c_\mathrm{p}}$ for $N_\mathrm{P}$ consecutive patches of the proteins (see Supplementary Information~\ref{sec:graph_construction} for details). $\mathbf{F}_\mathrm{P}$ and $\mathbf{F}_\mathrm{L}$ are combined to form a dense representation of the protein-ligand complex $\mathbf{F}\in \mathbb{R}^{(N_\mathrm{P}+N_\mathrm{L})\times (N_\mathrm{P}+N_\mathrm{L})  \times c_\mathrm{P}}$ which is further processed by the contact prediction and denoising modules (Figure~\ref{fig:fig2}b,e). 

\paragraph{Generating contact maps and pair representations}
\label{method:contact_map}
We define the contact map of a protein-ligand complex based on the pairwise proximities among all residues and selected frames (Methods, Distograms and contact maps). During model inference, the contact maps are modeled as the logits of a categorical posterior distribution;
the CPM auto-regressively refines a single realization of the contact map $\mathbf{L}$  using %
feedback from the previous iteration predictions. Specifically, the CPM samples a one-hot adjacency matrix from the last-iteration-predicted contact map and sums all sampled adjacency matrices $\{\mathbf{l}\}_{k=1}^{N_\mathrm{L}}$ as an additional signal passed to the neural network, until $k=N_\mathrm{L}$ when each ligand frame is assigned to a protein patch:
\begin{equation}
    \label{eq:contact_forward}
    \hat{\mathbf{L}}_{k+1} = \mathrm{CPM}( \sum_{r=1}^{k}{\mathbf{l}_r}, \mathbf{F}); \ \mathbf{l}_k = \mathrm{OneHot}(i_k, j_k); \ (i_k, j_k)\sim\mathrm{Categorical}_{N_\mathrm{P}\times N_\mathrm{L}}(\hat{\mathbf{L}}_{k}) %
\end{equation}
where $k\in\{0, \cdots, N_\mathrm{L}\}$. 
This auto-regressive sampling scheme accounts for the multi-modal nature of protein-ligand contact distributions and allows the model to be trained on diverse structural databases including complexes with multiple bound ligands (such as in substrate-cofactor interactions). Concurrently, the CPM maintains a sparsely connected graph representation of all protein residues and ligand atoms based on their chemical and spatial proximity; the sparse graph regularly communicates with the dense representation $\mathbf{F}$ through a cross-attention mechanism (Methods, CPM). The last iteration contact map $\hat{\mathbf{L}}_{N_\mathrm{L}}$ is passed back into the CPM network %
to produce the set of graph representations used by the ESDM for structure generation.

\paragraph{Equivariant diffusion for atomistic structure generation}
We leverage diffusion-based generative modeling~\cite{sohl2015deep,song2020score} to sample 3D molecular geometries from a learned statistical distribution. Diffusion  models introduce a forward stochastic differential equation (SDE) that diffuses data into a known distribution and a neural-network-parameterized reverse-time SDE that generates data by reverting the noising process. Building upon this formulation, NeuralPLexer generates binding complex structures from a reverse-time process (Figure~\ref{fig:fig2}d):
\begin{equation}
    \label{eq:reverse_sde}
    d\mathbf{Z}_t = [-\Theta \mathbf{Z}_t - \Sigma \Sigma^\top \mathbf{s}_{\phi}(\mathbf{Z}_t; t)]  dt + \Sigma d\mathbf{W}_t
\end{equation}
where $\mathbf{Z}_t$ denotes all 3D atomic coordinates of the binding complex structure at time $t$, and $\mathbf{s}_{\phi}$ is a network-predicted function called the \textit{score function}~\cite{song2020score} which drifts the geometry $\mathbf{Z}_t$ towards the target data distribution. We adopt $\Theta$ and $\Sigma$ matrices that involve anisotropic linear terms among chemically adjacent atoms (Figure~\ref{fig:fig2}c) with separated length-scale parameters such that %
global domain packing and local atomic structures are generated hierarchically. The diffusion processes are equivariant to translations and rotations, as detailed in Supplementary Information~\ref{sec:sde_se3}.
During inference, the atomic coordinates are first randomly initialized from a simple probability distribution that corresponds to the fully noised state of the forward-time SDE.
At each integrator step along the reverse-time SDE sampling~\eqref{eq:reverse_sde}, the ESDM utilizes a stereochemistry-aware graph transformer neural network (Figure~\ref{fig:fig2}e) to predict a set of denoised coordinates $\hat{\mathbf{Z}}_0$ which is then transformed to $\mathbf{s}_{\phi}$ %
(Methods, ESDM).
Moreover, we adopt a stochastic temperature-adjusted sampler (Langevin Simulated Annealing SDE, LSA-SDE) and introduce a correction scheme to improve consistency between integrator steps and reduce numerical discretization error (Methods, Temperature-annealed and continuity-corrected diffusion model sampling) which are found to improve accuracy.

\paragraph{Model training} To train NeuralPLexer, we introduce a dataset named PL2019-74k (Methods, Datasets and training summary) compiled from a diverse set of experimentally determined apo protein structures and protein-ligand complex structures from the Protein Data Bank (PDB), cross-referenced to annotations in other public datasets to systematically filter experimental artifacts. The loss function used for model training comprises a cross-entropy term for CPM outputs, a translation-rotation-invariant structure denoising term evaluated for global and binding site structures, and regularization terms to improve local distance-geometry quality and reduce structural violations. During model training, an MSA bootstrapping technique~\cite{del2022sampling} was applied to sample diverse template structures from AF2 as auxiliary model inputs complementary to alternative experimentally determined conformations of the same proteins retrieved from the PDB (Methods, Datasets and training summary). The training process in total took 16 days on 6 NVIDIA V100-32GB GPUs, which is overall substantially lower than the cost of training state-of-the-art protein structure prediction models (for example, 11 days on 128 TPU-v3 cores for training AF2~\cite{jumper2021highly}). 

\begin{table}
     \begin{small}
     \begin{center}
     \makebox[\textwidth][c]{
    \begin{tabular}{lcccccc|cccccc}
    
    \toprule
     &\multicolumn{6}{c}{Top-1 Ligand RMSD} & \multicolumn{5}{c}{Top-5 Ligand RMSD}\\
     &\multicolumn{3}{c}{Percentiles $\downarrow$} & \multicolumn{2}{c}{below threshold $\uparrow$} & GPU time (s)  & \multicolumn{3}{c}{Percentiles $\downarrow$} & \multicolumn{2}{c}{below threshold $\uparrow$}\\

    \textbf{Methods} & 25th & 50th & 75th & 5 \AA{}  &  2 \AA{} & & 25th & 50th & 75th & 5 \AA{}  &  2 \AA{} \\
    \midrule
    P2Rank+GNINA~\cite{corso2022diffdock} & 1.7 & 5.5 & 15.9 & 47.8 & 28.8 & 127 & 1.7 & 5.5 & 15.9 & 47.8 & 28.8  \\ 
    DiffDock~\cite{corso2022diffdock} & 1.4 & 3.3 & 7.3 & 63.2 & 38.2 & 40 & 1.2 & 2.4 & 5.0 & 75.5 & 44.7 \\ 
    PointSite+Uni-dock~\cite{yu2023deep} & - & 5.5 & - & - & 32.1 & - & - & 2.5 & - & - & 46.1  \\ 
    DiffDock+Uni-dock~\cite{yu2023deep} & - & 4.1 & - & - & 38.9 & - & - & 1.9 & - & - & 51.1  \\ 
    NeuralPLexer (ours) & \textbf{1.3} &  \textbf{2.8} & \textbf{5.9} & \textbf{69.7} & \textbf{39.5} & \textbf{2.1} & 1.1 & 2.3 & 4.6 & 77.0 & 47.0 \\ 
    \bottomrule
    \end{tabular}}
    \end{center}
    \end{small}
    \caption{Summary of performance statistics for rigid receptor blind protein-ligand docking on the PDBBind2020 benchmark. Statistics for the baseline methods are obtained from the sources referenced in their respective rows. Note that the top-5 statistics reported by baseline methods may depend on their pose selection schemes; the NeuralPLexer top-5 results are computed using 5 randomly sampled poses per protein-ligand pair, thus should be considered a success rate lower bound.}
    \label{tab:table1}
\end{table}

\begin{figure}
    \centering
    \includegraphics[width=\textwidth]{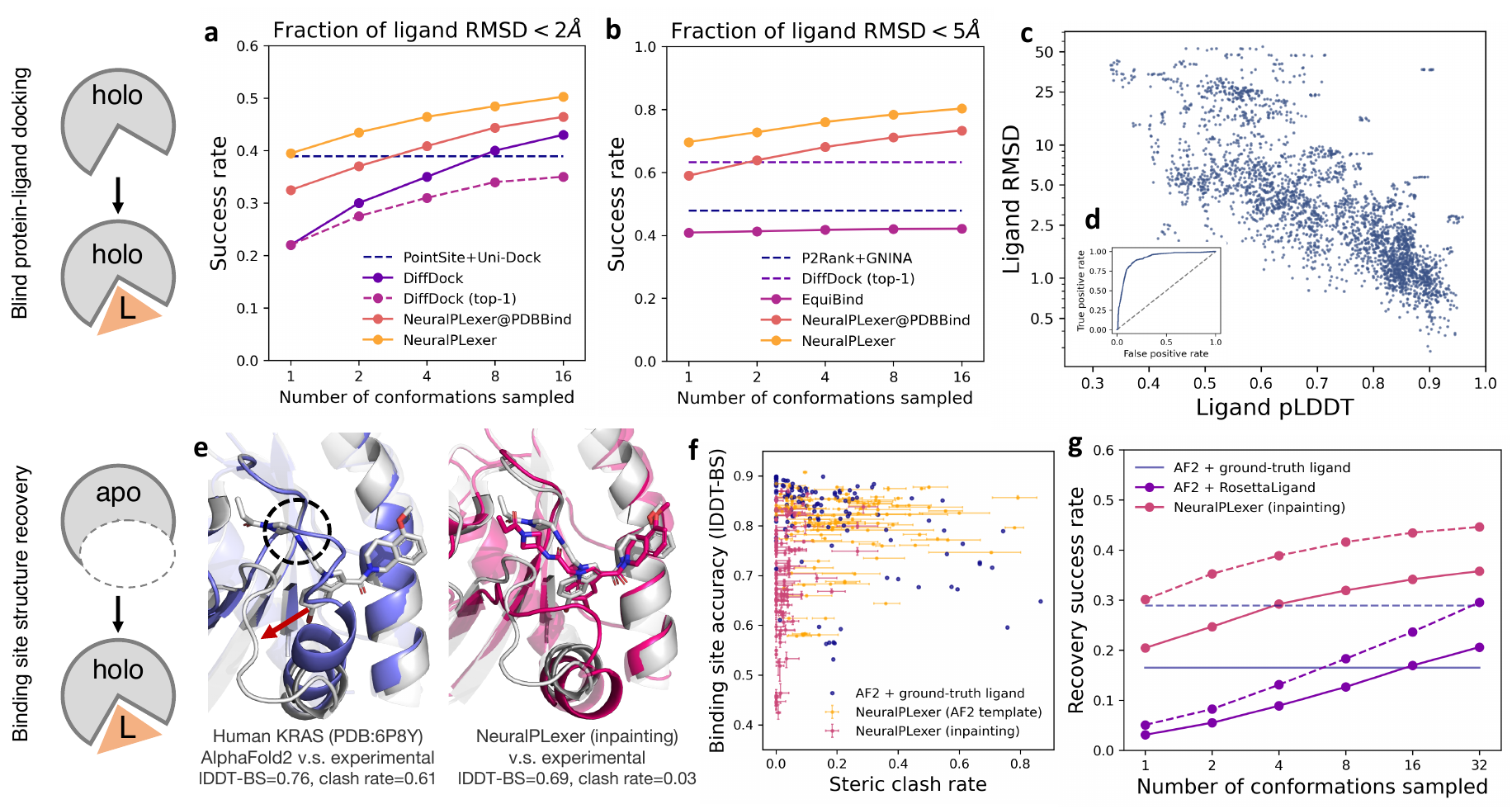}
    \caption{Model performance on benchmarking problems. (a-d) Rigid backbone blind protein-ligand docking. The cumulative fraction of predictions with ligand RMSD below (a) \SI{2}{\angstrom} and (b) \SI{5}{\angstrom} over the test dataset are plotted against the number of ligand poses sampled per protein-ligand pair. (c) Assessing the fidelity of model-assigned confidence estimations. The ligand RMSDs are plotted against the model-assigned pLDDT score averaged over all ligand atoms (ligand pLDDT). (d) The precision-recall curve is evaluated based on the sampled structures ranked by the ligand pLDDT score, with the binary value of ligand RMSD < \SI{2}{\angstrom} being treated as the class label. 
    (e-g) Flexible binding site structure recovery. (e) Visualization of prediction results on a test set example (PDB:6P8Y) near the structure recovery accuracy cutoff considered in this work. NeuralPLexer generates the binding site protein-ligand structure consistent with experiments, while directly aligning the AF2 prediction to ground-truth complex results in steric clashes (dashed circle). The red arrow indicates the qualitative conformational change from AF2 to the experimental bound-state structure. (f) Summary of binding site accuracy (measured by the lDDT-BS score) and ligand clash rate over the test dataset. 32 conformations are sampled for each protein-ligand pair; dots indicate the median value and error bars indicate 25th and 75th percentiles. (g) The structure recovery accuracy is compared to baseline methods. Solid lines correspond to recovery rates evaluated based on strict cutoffs (lDDT-BS > 0.7, ligand RMSD < $\SI{2.0}{\angstrom}$, clash rate = 0.0), while the dashed lines correspond to relaxed cutoffs (lDDT-BS > 0.5, ligand RMSD < $\SI{2.5}{\angstrom}$, clash rate < 0.05) %
    with clash rate cutoff matching the 95\% percentile of experimental structure statistics.}
    \label{fig:fig3}
\end{figure}

\subsection*{Accurate ligand and binding pocket structure modeling}

We evaluate NeuralPLexer on two benchmarking tasks that are simplified sub-problems for protein-ligand complex structure prediction, yet they are still considered challenging problems for conventional molecular docking workflows. Specifically, we first benchmark the method on \textit{blind protein-ligand docking} where the ground-truth receptor protein structure is given as input, and the ligand coordinates are predicted without any pre-specified binding site information. %
Model predictions are collected on the PDBBind2020~\cite{wang2005pdbbind} dataset, a community-wide recognized benchmark for protein-ligand docking.
We compare our results to alternative computational algorithms such as strategies combining binding site prediction models (such as PointSite~\cite{yan2022pointsite} and P2Rank~\cite{krivak2018p2rank}) and conventional search-and-scoring-based docking algorithms (such as GNINA~\cite{mcnutt2021gnina} and Uni-Dock~\cite{yu2022uni}) for blind docking, as well as a recent generative-model-based method DiffDock~\cite{corso2022diffdock} that is state-of-the-art on this benchmark. 
Using the dataset split in Ref.~\cite{stark2022equibind}, we assess the results from a NeuralPLexer model that was directly trained on the PDBBind2020 training set (Figure~\ref{fig:fig3}, NeuralPLexer@PDBBind) from randomly initialized parameters as well as a model that is fine-tuned from the checkpoint trained for end-to-end structure prediction on PL2019-74k (Figure~\ref{fig:fig3}, NeuralPLexer).
As shown in Table~\ref{tab:table1} and Figure~\ref{fig:fig3}a-c, NeuralPLexer-generated ligand poses achieve improved geometrical accuracy against reference methods, as consistently confirmed by the highest fraction of predictions with a ligand-heavy atom root-mean-square-deviation (RMSD) below the \SI{2.0}{\angstrom} and the \SI{5.0}{\angstrom} thresholds; pre-training the model on end-to-end structure prediction additionally improved success rate by around 20\% on both the RMSD<\SI{2.0}{\angstrom} and the RMSD<\SI{5.0}{\angstrom} criteria, suggesting that learning on tasks related to protein folding prediction produces model representations that are better suited to identify functional ligand binding sites.
To study how well the generated ligand structure distribution covers the ground truth pose, in Figure~\ref{fig:fig3}a-b we also plot the fraction of targets for which a successful prediction can be identified against the number of docking trials per target (x-axis). Remarkably, 39.5\% of NeuralPLexer predictions are accurate to within an RMSD of \SI{2.0}{\angstrom} despite sampling only 1 ligand pose per target, representing a 78\% improvement compared to the best competing method, DiffDock, at this limit. This result highlights the effectiveness of incorporating contextual information through the form of contact maps to localize the space of structural hypotheses, 
and implies the potential to accelerate virtual screening while easing the need for designing scoring functions.
Last, we evaluate the relationship between the ligand pose prediction accuracy against the model-assigned confidence estimations; as shown in Figure~\ref{fig:fig3}c-d, the predicted pLDDT scores averaged over ligand atoms are well correlated with the true ligand RMSD, and 80\% of the predicted structures with RMSD<\SI{2.0}{\angstrom} can be identified by ranking the structures using ligand pLDDT, with a low false positive rate of 11.6\%.

\begin{figure}
    \centering
    \includegraphics[width=\textwidth]{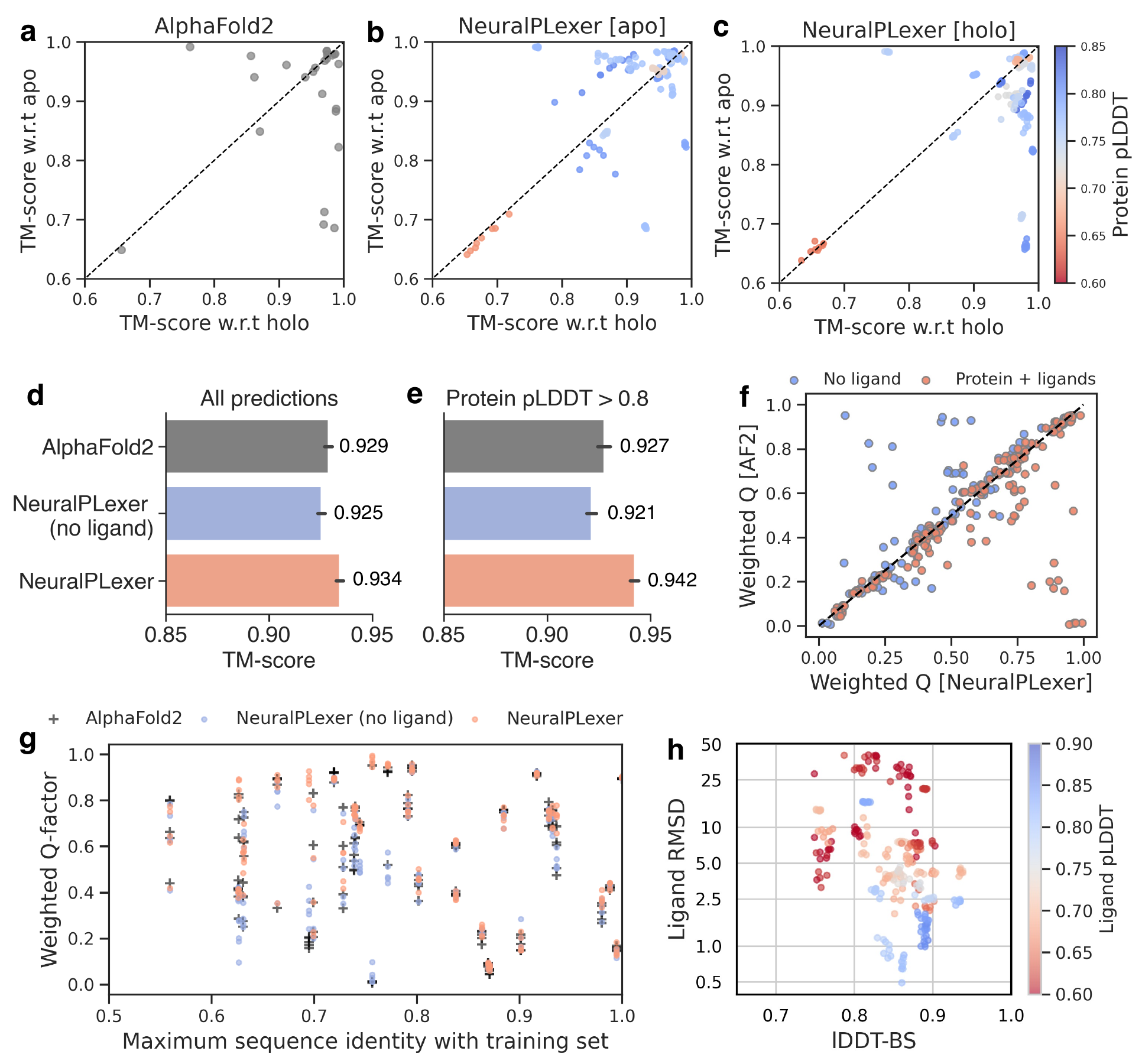}
    \caption{Model predictions for contrasting apo-holo pairs from the PocketMiner dataset. (a-c) The TM-score with respect to apo and holo experimental reference structures are plotted for (a) AlphaFold2 (AF2) structures (b) NeuralPLexer-sampled apo structures, and (c) NeuralPLexer-sampled holo structures. For NeuralPLexer-sampled structures, the model-assigned pLDDT scores averaged among all protein residues are indicated by the dot colors. (d-e) TM-scores for baseline methods and NeuralPLexer averaged against all samples and the subset of targets for which any NeuralPLexer-predicted structure is of protein pLDDT=0.8 or higher. Error bars indicate the standard error of the mean calculated from six sets of predictions using independent random seeds and different AF2 model checkpoints to obtain the initial template. (f) The weighted Q-factor metric for conformational change prediction accuracy (see Methods, Weighted Q-factor) is plotted between NeuralPLexer predictions and AF2 predictions on all ligand-bound holo structures. Grey dots correspond to NeuralPLexer predictions in the absence of ligand inputs. (g) The weighted Q-factors for predicted ligand-bound holo structures are plotted against the maximum sequence similarity of each target to samples in the training dataset. (h) Assessing the fidelity of NeuralPLexer internal confidence predictions. The ligand-binding-site accuracy as measured by lDDT-BS is plotted against the ligand RMSD for all NeuralPLexer predictions; the model-assigned pLDDT scores averaged among all ligand atoms are indicated by the dot colors.}
    \label{fig:fig4}
\end{figure}

We then apply NeuralPLexer to perform \textit{binding site structure recovery}, which can be interpreted as %
a principled approach to model highly flexible binding pockets. In particular, given a ligand-free receptor protein conformation and the binding site centroid coordinate, we adopt a fine-tuned NeuralPLexer model to jointly predict the structure for a cropped spherical region within $\SI{6.0}{\angstrom}$ of any ligand atom by inpainting all the amino acid and ligand atomic coordinates conditioning on the uncropped parts of the receptor backbone. 
The binding pocket structure accuracy relative to the reference experimental structures is measured by the lDDT-BS, the all-atom Local Distance Difference Test~\cite{mariani2013lddt} metric averaged for residues within $\SI{4.0}{\angstrom}$ of any ligand atom %
consistent with CAMEO~\cite{robin2021continuous}. 
Input backbones are obtained using template-free AF2 predictions of 154 selected chains with TM-score>0.8 (that is, high backbone accuracy) but with lDDT-BS<0.9 out of the PDBBind2020 test set, a subset representing cases where AF2 correctly predicts the global protein folding but fails to reproduce the exact bound-state binding site structure. %
We first assessed the fidelity of these initial AF2 structures for ligand docking by computing the steric clash rate between the ligand and the receptor, defined as the fraction of ligand heavy atoms with a Lennard-Jones energy > 100 kcal/mol using UFF~\cite{rappe1992uff} parameters. The stringency of this criterion is reflected by the observation that 9\% of the experimental structures are classified as clash-containing because of experimental errors and structures with cross-linking.
Out of the 154 AF2-predicted receptor structures, we find that only 18\% of structures contain no steric clash with the ligand when directly aligned to the reference experimental binding complex, which represents a success rate upper bound for template-based ligand modeling for such targets if using AF2 structures as references. In contrast, as shown in Figure~\ref{fig:fig3}f-g (NeuralPLexer-inpainting), our modeling strategy is able to accurately dock ligands to the specified sites while retaining a low clash rate and competitive binding pocket structure prediction accuracy which is evident from an improved successful recovery rate of up to 35\% (and up to 46\% on a relaxed criterion). 
The overall good all-atom binding site accuracy on all prediction targets as measured by lDDT-BS suggests that the improvement in ligand geometric accuracy and clash rate is not an effect of over-aggressively relaxing the binding site residues. 
As a reference, we perform inference using the entire AF2 structure as input template (Figure~\ref{fig:fig3}f, NeuralPLexer (AF2Template)) instead of only using non-binding-site residues; despite a similar ligand median RMSD of \SI{3.34}{\angstrom} and a higher average lDDT-BS score of 0.82 compared to NeuralPLexer-inpainting (RMSD=\SI{3.51}{\angstrom}, lDDT-BS=0.71), NeuralPLexer (AF2Template) resulted in a substantially higher fraction of clash rate (0.22 on average) which implies the deficiency of these AF2 models for standard rigid-protein molecular docking.

Figure~\ref{fig:fig3}e shows a prediction example on human KRAS\textsuperscript{G12C} with a cysteine-reactive covalent inhibitor~\cite{shin2019discovery} (PDB:6P8Y) for which the opening of Switch-II pocket~\cite{moore2020ras} harbors an unconventional druggable site. While the AF2 prediction for this target recapitulates the native-like closed pocket with a severe steric clash with the crystal structure ligand, the NeuralPLexer-inpainted binding site successfully predicted an open-like conformation with a ligand RMSD of \SI{2.08}{\angstrom}. Although NeuralPLexer qualitatively improved the binding site conformation model for this target, we still observed a slight drop in the measured lDDT-BS score from 0.76 of AF2 to 0.69. This observation suggests that the combined metric used in Figure~\ref{fig:fig3}g is a more appropriate criterion to assess binding site structure prediction fidelity, as the simultaneous satisfaction of low ligand RMSD and clash rate implies a correct pocket shape which may not be discerned by distance-based metrics such as lDDT.
We have also compared our quantitative results against a baseline method RosettaLigand~\cite{davis2009rosettaligand}, an algorithm designed for ligand docking with fast energy-based flexible receptor relaxation (Figure~\ref{fig:fig3}g). We confirm that NeuralPLexer achieves consistently higher structure recovery rates without applying any posthoc filtering, even though RosettaLigand explicitly used the entire AF2 structure as the initial guess for relaxation while NeuralPLexer only used the binding-site-cropped AF2 scaffold as input. The capability of inpainting cropped structures without template information shows immediate applicability for de novo ligand-binding protein design~\cite{polizzi2020defined}, since the NeuralPLexer network can be viably adapted to jointly generate binding site backbones and sequences in the absence of protein sequence inputs.

\begin{figure}
    \centering
    \includegraphics[width=\textwidth]{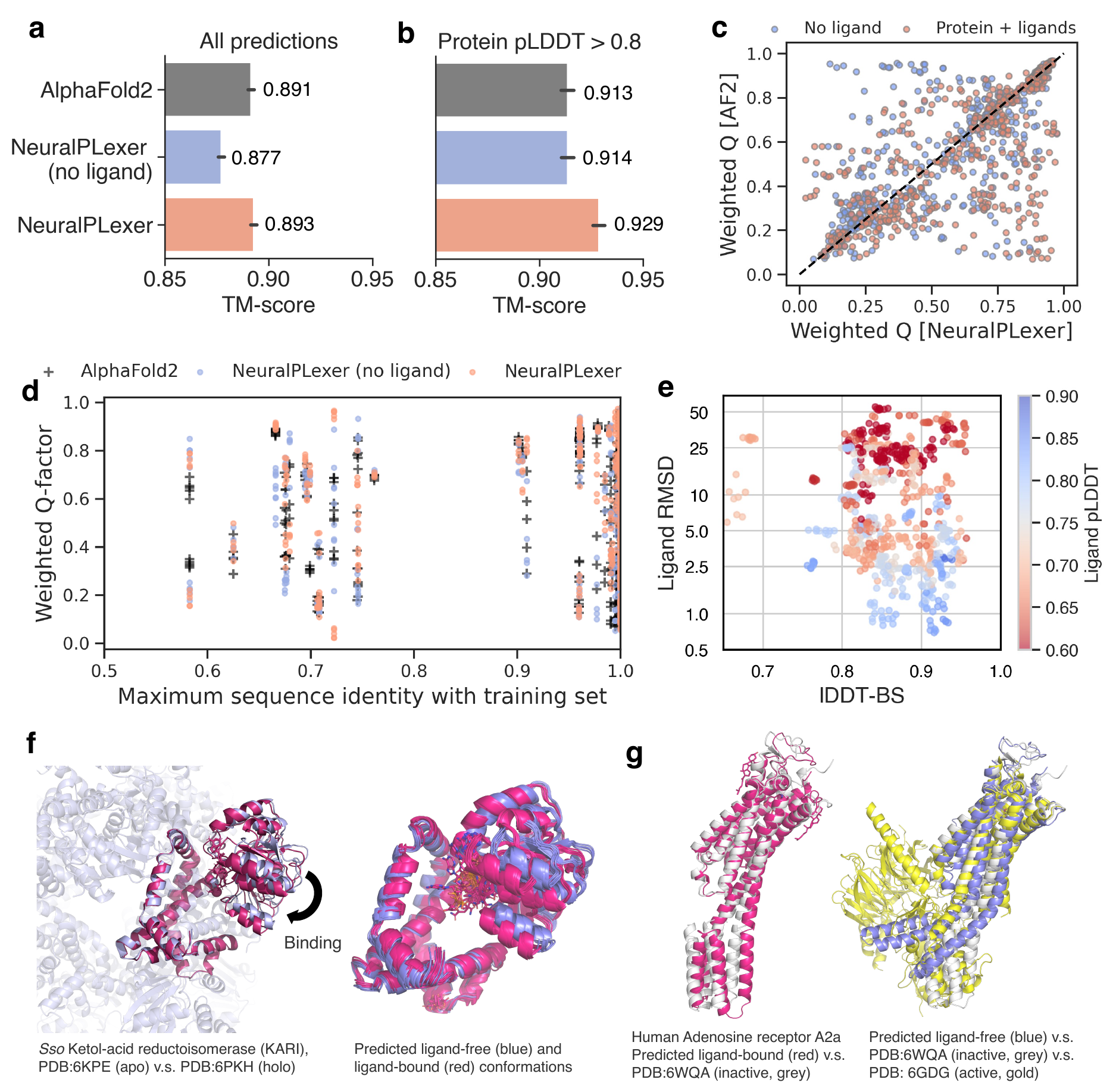}
    \caption{Model predictions for recently determined structures. %
    (a-b) TM-scores for baseline methods and NeuralPLexer averaged against all samples and the subset of targets for which any NeuralPLexer-predicted structure is of protein pLDDT=0.8 or higher. Error bars indicate the standard error of the mean calculated from six sets of predictions using independent random seeds and different AF2 model checkpoints to obtain the initial template. (c) The weighted Q-factor metric for conformational change prediction accuracy (see Methods, Weighted Q-factor) is plotted between NeuralPLexer predictions and AF2 predictions on all ligand-bound holo structures. Grey dots correspond to NeuralPLexer predictions in the absence of ligand inputs. (d) The weighted Q-factors for predicted ligand-bound holo structures are plotted against the maximum sequence similarity of each target to samples in the training dataset. (e) Assessing the fidelity of NeuralPLexer internal confidence predictions. The ligand-binding-site accuracy as measured by lDDT-BS is plotted against the ligand RMSD for all NeuralPLexer predictions; the model-assigned pLDDT scores averaged among all ligand atoms are indicated by the dot colors. (f-g) Model predictions on representative targets (PDB:6KPE, PDB:6PKH, PDB:6WQA) suggest structural elements for protein self-assembly and plausible models for enzyme catalysis and the activation of membrane receptors.}
    \label{fig:fig5}
\end{figure}

\subsection*{End-to-end structure prediction for ligand-binding proteins} 

Having established a new mark on benchmarking problems, we then sought to evaluate the performance of NeuralPLexer on predicting the structures of ligand-binding proteins %
that exhibit large conformational variability. We first assessed NeuralPLexer-predicted structures on 33 contrasting apo-holo pair systems from the PocketMiner~\cite{meller2022predicting} dataset, which was held out from our training datasets based on their protein UniProt ID such that no structures with identical protein sequence is observed during training. To examine the overall prediction accuracy of such ligand-binding systems, we compared the NeuralPLexer predictions against the best-performing protein structure prediction algorithm AF2. 
Figure~\ref{fig:fig4}d and Table~\ref{table:pred_pocketminer} 
summarizes the prediction accuracy statistics as evaluated by TM-score~\cite{zhang2005tm}, on which NeuralPLexer exhibits the highest performance.
We delineate the underlying factors that contributed to the improved prediction accuracy by performing pairwise structure comparisons against their corresponding experimental apo and holo reference structures; as shown in Figure~\ref{fig:fig4}b, the protein structure ensembles sampled from NeuralPLexer in the absence of ligand inputs are overall diverse and more similar to the experimental apo reference structures, while the protein parts of the binding complexes obtained with both protein and ligand graph inputs are well-aligned with the experimental holo structures. Instead, AF2 produces a mixture of structures for which no consistent trend of relative similarity to apo or holo states was observed, which leads to a decrease in TM-score when averaged against all apo and holo samples. In summary, we observe an average TM-score=0.934 for structures sampled from NeuralPLexer, representing a statistically-significant improvement of TM-score=0.929 from AF2 (p=0.03) and TM-score=0.925 from a ligand-free baseline model (p=0.0004) for which we ablate all ligand graph inputs from NeuralPLexer to only sample apo structures (Figure~\ref{fig:fig4}d, NeuralPLexer (no ligand)).
The moderate decrease of structure accuracy from NeuralPLexer (no ligand) compared to AF2 additionally supports our hypothesis that the state selectivity between apo and holo structures is the key factor to the improvement in overall prediction accuracy, even for cases where the overall protein structure generation quality is lower than AF2 due to the lightweight nature of the network and training data. 

Apart from standard C$\alpha$-based metrics such as TM-score, to further elucidate the all-atom model prediction accuracy for regions that undergo major structural changes upon ligand binding, we introduce a weighted version of the Q-factor from Best and Hummer~\cite{best2013native} based on the inter-residue native contacts that are not conserved between two distinct-state reference structures (Methods, Weighted Q factor). As shown in Figure~\ref{fig:fig4}f, we observe a stark differentiation between NeuralPLexer (average Q-factor=0.608) and the ligand-free baseline model (average Q-factor=0.501), while 16\% of the predicted structures are found to qualitatively improve against AF2 models (average Q-factor=0.538) by %
0.1 or more.
These results clearly indicate that NeuralPLexer is able to selectively sample ligand-free and ligand-bound protein states on targets that are challenging to predict using state-of-the-art methods. %
The prediction accuracy as measured by Q-factor is also consistent among targets with different sequence similarities to the training dataset (Figure~\ref{fig:fig4}f), suggesting that the model is able to generalize beyond targets for which structures of close homologs are available. In addition to the overall prediction accuracy, we find that the NeuralPLexer-assigned pLDDT score can effectively identify more accurately predicted structures for such ligand-binding proteins, as manifested by the increase from TM-score=0.92 of AF2 to 0.95 of NeuralPLexer when only evaluated on the subset at which model is most confident (average protein residue pLDDT > 0.8).

To determine the accuracy of NeuralPLexer on novel protein-ligand systems, we then assess the predicted structures on 118 recently resolved protein and protein-ligand complexes from structures that are deposited to the PDB since January 2019 (Figure~\ref{fig:fig5}, Table~\ref{table:pred_recent}). These targets are selected %
based on the criterion that a reference experimental structure of the same protein with sequence identity >98\% %
and backbone RMSD>\SI{2.0}{\angstrom} can be found from the PDB. Despite that 48 out of 98 holo structures in the dataset have homologs in the PDB with 100\% sequence identity, the large conformational variability of these samples still renders it nontrivial to predict their structures using standard approaches. The prediction difficulty of these targets is evident from an overall poor Q-factor of 0.541 for AlphaFold2 predictions, since a contact map with random binary values would produce a Q-factor of 0.5. A similar result of average Q-factor=0.539 is observed for ligand-free NeuralPLexer predictions. In contrast, NeuralPLexer with ligand input improves the average Q-factor from 0.541 to 0.603. Out of the 50 holo targets for which identical-sequence homologs are not included within training (Figure~\ref{fig:fig5}d), the NeuralPLexer prediction on 13 targets shows a significant increase in Q-factor by 0.1 or more, as opposed to 7 targets for the ligand-free baseline. Overall, on this dataset NeuralPLexer improves the average protein structure prediction accuracy by a TM-score increase from 0.877 that of NeuralPLexer (no ligand) to 0.893 which matches the accuracy of AF2 (average TM-score=0.891), and outperforms AF2 by an improvement from average TM-score=0.913 to TM-score=0.929 for the subset with protein pLDDT>0.8.
On both test datasets, more than 70\% predictions that are of ligand RMSD<\SI{2.5}{\angstrom} and lDDT-BS>0.8 can be distinguished from others by a criterion of ligand pLDDT score > 0.8 (Figure~\ref{fig:fig4}e, Figure~\ref{fig:fig5}e), which confirms that the model-assigned confidence scores also provide a consistent metric to identify highly accurate predictions for both ligand and binding site residues.

\subsection*{Interpreting allostery and catalytic mechanisms from predicted structures} %

Furthermore, an examination of model predictions for recently determined structures reveals that the conformational variations among the predicted structures can be leveraged to gain insight into protein function and their interactions. On a ketol-acid reductoisomerase target (PDB:6KPE, Figure~\ref{fig:fig5}f) whose catalytic mechanism was recently characterized through cryo-EM experiment~\cite{chen2019temperature}, by comparing the predicted ligand-bound and ligand-free ensembles we find that NeuralPLexer accurately captures the closure motion of the N-subdomain upon NADH and inhibitor binding. Apart from the structural domains with high conformational mobility, the protein was found to spontaneously form a dodecamer through the self-assembly of the C-subdomain which is also consistent with the low structural fluctuation near the observed protein-protein interface among the NeuralPLexer-sampled conformations.
The predicted structures may also help the identification of key structural elements that are critical for protein activation and deactivation, especially for complex systems where experimental structure determination is challenging such as in G protein-coupled receptors (GPCRs). For example, Figure~\ref{fig:fig5}c shows representative predicted structures for the human Adenosine receptor A2a - Cytochrome b562 chimera (PDB:6WQA)  on which the predicted antagonist-bound structure agrees well with the structure determined by XFEL experiments~\cite{lee2020harnessing} with a TM-score=0.86. In contrast, the predicted ligand-free structures contain snapshots with significant bending near the terminal of helices TM5 and TM6; although no apo experimental structure has been determined for this target, structure alignment with the G-protein-bound active conformation~\cite{garcia2018cryo} (PDB:6GDG) suggests that the sampled ligand-free structures are plausible intermediate states that may be attributed to constitutive activity in the absence of ligands~\cite{bertheleme2013loss}. 
The ability of NeuralPLexer to generate faithful structural ensembles may enable a powerful tool for generating hypotheses to help unravel various molecular mechanisms related to allosteric regulation and enzyme catalysis.

\section*{Discussions}

Determining the structures and functional conformational changes of protein-ligand complexes at the proteome scale is a grand-challenge problem. We have demonstrated a key step towards solving this problem by showing that NeuralPLexer, a generative deep learning approach that systematically incorporates small molecule information and biophysical inductive bias, achieves consistent improvement against state-of-the-art molecular docking and protein structure prediction methods. NeuralPLexer is capable of generating structures within seconds on a standard GPU once all auxiliary features are gathered, thereby enabling the exploration of protein-ligand interactions on the vast space of sequence and chemical spaces by leveraging rapidly evolving databases of computationally modeled protein structures~\cite{tunyasuvunakool2021highly,lin2023evolutionary} and chemical substances~\cite{wishart2022hmdb,irwin2005zinc}. 

NeuralPLexer is end-to-end differentiable and thus is also well-suited for applications related to ligand and protein design. A closed-loop combination of NeuralPLexer and recent advances in differentiable generative models for protein sequences~\cite{bepler2021learning,rives2021biological,elnaggar2021prottrans}, molecular graphs~\cite{zang2020moflow,fu2021differentiable}, and structure-based bioactivity models holds the promise of accelerating the process of sequence and compound prioritization over conventional screening-based strategies. We also anticipate that NeuralPLexer can be directly applied to accelerate various physical simulation studies on protein-ligand interactions, such as guiding the design and optimization of collective variables in enhanced-sampling molecular dynamics simulations~\cite{zhao2021enhanced}. 

As a data-driven approach, NeuralPLexer is both generalizable and amenable to continuous improvement through the integration of better experimental and bioinformatic data. Improvements in the curation of training and benchmark datasets from a broader community could potentially enable a more systematic analysis of protein families with no experimentally determined homologs, as well as extending the method to more challenging systems such as post-translational modifications and multi-state large heteromeric protein complexes~\cite{plested2016structural}. %
Moreover, refining the model on data such as high-resolution nuclear magnetic resonance (NMR) and molecular dynamics data can enable it to capture protein structure under physiological conditions beyond the distribution of crystal-like structures.
A related promising direction to improve the methodology is to incorporate highly accessible auxiliary data related to protein-compound interactions such as binding affinities~\cite{liu2007bindingdb} and high-throughput mass spectrometry signals~\cite{piazza2018map}, %
given that the incorporation of large-scale protein sequencing data in the form of MSAs has already been demonstrated a crucial component to achieve transferable protein structure prediction~\cite{ovchinnikov2017protein,jumper2021highly,baek2021accurate}. Our study has provided a general computational framework for the exploration of these directions, paving the way toward the rapid and accurate protein-ligand complex structure prediction to facilitate structure biology, drug discovery, and protein engineering.

\clearpage

\section*{Methods}

\subsection*{Neural network architecture overview}

\paragraph{Molecular graph encoding.} We introduce Molecular Heat Transformer (MHT), an attention-based neural network architecture for molecule inputs with learned edge encodings to infer inter-atomic geometrical correlations based on their graph-topological representations. The encoder is composed of 8 MHT blocks, where a single MHT block comprises a multi-head self-attention block with edge embeddings added to the computation of attention weights (MHAwithEdgeBias, Supplementary Information~\ref{sec:chemfeat}), a pair update block, and a node update block. As a unique component of MHT, the pair update block updates the pair representations between the set of atomic nodes and frame nodes through discretized heat-kernel transformations~\cite{kondor2002diffusion}, such that they are capable of incorporating the molecular graph chirality and are able to explicitly encode long-range dependencies beyond neighboring nodes on the molecular graph. 

To train the MHT model in NeuralPLexer, we designed a specialized algorithm to align the learned molecular representations with both %
3D molecular structure and %
bioactivity information. In particular, we introduced a hybridized loss function that comprises a mixture density network term to fit the 3D distance-angle distributions to the reference molecular geometries for all edges between the atom and frame nodes, a global denoising term to recover the molecular geometry from perturbed geometries sampled from a forward-time SDE, a supervised loss term to predict the bioactivity labels following a graph pooling operation, as well as a masked language modeling loss. The featurization details, MHT architecture, related datasets, and additional training details are described in Supplemental Information~\ref{sec:small_mol_pretraining}.

\paragraph{Protein-ligand graph representation and the contact prediction module (CPM).} As detailed in Supplementation Information~\ref{sec:graph_construction}, we adopt a residue-scale graph representation combining sparse edges determined based on the input noisy protein geometry $\mathbf{x}_t$ and densely connected edges associated with pair representations $\mathbf{F}$ for a selected subset of protein and ligand nodes ("anchor nodes"). The protein anchor nodes are selected by first sequentially segmenting the input protein sequence into $N_\mathrm{P}$ equal-sequence-length patches, and then sampling one unique backbone frame for each protein patch; the $N_\mathrm{L}$ ligand anchor nodes are uniformly sampled from all $N_\mathrm{frame}$ frame nodes of the ligand molecular graphs. Protein node features are initialized from the PLM embeddings of the input sequence (or the concatenation of PLM embeddings of all sequences in $\{\mathbf{s}\}$, when the input is a protein complex); the intra-protein edge features are initialized via a geometrical encoding of the relative backbone frame distances and orientations between the source and destination nodes. When the template protein structure input is present, local encodings of side-chain internal coordinates and relative geometrical encodings of the template backbone frames are respectively added to node and edge features. 
The CPM (Figure~\ref{fig:fig2}b, Supplementation Information~\ref{sec:CPM}) enables expressive and scalable processing of the residue-scale graph, featuring a linear scaling with respect to the number of nodes and a constant overhead. The neural network architecture %
comprises a graph attention layer to update all node and edge representations, a triangular attention layer to update the densely-connected edge representations while accounting for edge-edge interactions, and a cross-attention layer to synchronize information between the sparse and the dense components of the graph. 

\paragraph{The equivariant structure denoising module (ESDM).}
The architecture of ESDM (Figure~\ref{fig:fig2}e, Supplementary Information~\ref{sec:ESDM}) is inspired by prior works on 3D graph and attentional neural networks for point clouds~\cite{satorras2021n,brandstetter2021geometric}, rigid-body simulations~\cite{li2018learning} and biopolymer representation learning~\cite{jing2020learning,jumper2021highly,shen2022e2efold,anand2022protein}. In ESDM, each node is associated with a stack of standard scalar features $\mathbf{f}_\mathrm{s}\in\mathbb{R}^{c}$ and cartesian vector features $\mathbf{f}_{\mathrm{v}}\in\mathbb{R}^{3\times c}$ representing the displacements of a virtual point set relative to the node's coordinate $\mathbf{t}\in \mathbb{R}^3$. A rotation matrix $\mathbf{R}\in\mathrm{SO(3)}$ is additionally attached to each rigid-body (i.e., frame) node. Geometry-aware messages are synchronously propagated among all nodes by encoding the pairwise distances among virtual point sets into graph transformer blocks (PointSetAttentionwithEdgeBias, Algorithm~\ref{alg:PointSetAttentionwithEdgeBias}).
Explicit non-linear transformation on vector features $\mathbf{f}_{\mathrm{v}}$ is solely performed on rigid-body nodes through a coordinate-frame-inversion mechanism (LocalUpdateUsingReferenceRotation, Algorithm~\ref{alg:LocalUpdateUsingReferenceRotation}), such that the node update blocks are sufficiently expressive without sacrificing equivariance or computational efficiency. 
On the contrary, 3D coordinates are solely updated for atomic nodes (PredictDrift, Algorithm~\ref{alg:PredictDrift}) while the rigid-body frames $(\mathbf{t}, \mathbf{R})$ are passively reconstructed according to the updated atomic coordinates, circumventing numerical issues regarding fitting quaternion or axis-angle variables when manipulating rigid-body objects. 
The nontrivial actions of an inversion operation on rigid-body nodes ensure that ESDM captures the correct chiral-symmetry-breaking behavior that adheres to the molecular stereochemistry constraints.

\subsection*{Distograms and contact maps} 
Given the atomic coordinates of a binding complex, we define the ground-truth contact map as 
\begin{equation}
    \label{eq:ContactMap}
    L_{AJ}= \mathrm{max}(1 - \frac{\lVert \mathbf{x}_A - \mathbf{y}_J \rVert}{D_0},\, \varepsilon)
\end{equation}
where $\mathbf{x}_A$ stands for the centroid coordinate of amino acid residue $A$, and $\mathbf{y}_J$ stands for the center-atom coordinates of a selected ligand frame $J$; we use $D_0=\SI{8.0}{\angstrom}$ and $\varepsilon=10^{-6}$ for all results reported in this study. At each auto-regressive contact map refinement iteration, the CPM network outputs histograms of pairwise distances ("distograms") $\hat{P}(d_{AJ} | \{\mathbf{s}\}, \{G\}, \sum_{r=1}^{k} \mathbf{l}_r)$ between all protein residues $A\in\{1, 2, \cdots, N_\mathrm{res}\}$ and ligand frames $J\in\{1, 2, \cdots, N_\mathrm{frame}\}$. The distograms are then used to estimate the predicted contact map $\hat{L}$
\begin{equation}
    \label{eq:DistogramToContactMap}
    (\hat{L})_{AJ} = \sum^{N_\mathrm{bins}} \hat{P}(d_{AJ} | \{\mathbf{s}\}, \{G\}, \sum_{r=1}^{k} \mathbf{l}_r) \cdot \mathrm{max}(1 - \frac{d_{AJ}}{D_0}, \, \varepsilon), \quad d_{AJ} \in [d_\mathrm{min}, d_\mathrm{min} + \frac{d_\mathrm{max}-d_\mathrm{min}}{N_\mathrm{bins}-1}, \cdots, d_\mathrm{max}]
\end{equation}
we use $N_\mathrm{bins}=32$, $d_\mathrm{min}=\SI{2.0}{\angstrom}$, and $d_\mathrm{max}=\SI{22.0}{\angstrom}$ for the distograms. The contact map $\hat{L}$ is further coarse-grained into the respective protein patches before sampling the block-adjacency matrices (Eq.~\eqref{eq:contact_forward}) to be recycled by the CPM network.

\subsection*{Diffusion SDE details}

To motivate the design principles for our diffusion-based biomolecular structure generator, we first consider a general class of linear SDEs known as the multivariate Ornstein–Uhlenbeck process~\cite{meucci2009review} for point cloud $\mathbf{Z}\in \mathbb{R}^{N\times 3}$: %
\begin{equation}
    d\mathbf{Z}_t = -\Theta \mathbf{Z}_t dt + \Sigma d\mathbf{W}_t
\end{equation}
where $\Theta\in\mathbb{R}^{N\times N}$ is an invertible matrix of affine drift coefficients, $\Sigma\in\mathbb{R}^{N\times N}$ represents the factorized covariance matrix, and $\mathbf{W}_t$ is a standard $3N$-dimensional Wiener process. The forward noising SDEs used in standard diffusion models~\cite{song2019generative,ho2020denoising} %
can be recovered by setting $\Theta=\theta \mathbf{I}$, %
converging to an isotropic Gaussian prior distribution at the $t\rightarrow \infty$ (often expressed as $t\rightarrow 1$ with reparameterized $t$~\cite{karras2022elucidating}) limit. %
Instead, we introduce a multivariate SDE containing data-dependent drift and covariance matrix $\Theta$ %
of a factorized form; %
\begin{equation}
    d\mathbf{Z}_t = -(\mathbf{U} \bm{\lambda} \mathbf{U}^{-1} ) \mathbf{Z}_t dt + (\sigma \mathbf{U}) d\mathbf{W}_t
\end{equation}
where $\mathbf{Z}=[\mathbf{x}_\mathrm{C\alpha}, \mathbf{x}_\mathrm{nonC\alpha}, \mathbf{y}]^\top$ according to the following symbolic convention
\begin{itemize}
    \item $\mathbf{x}_\mathrm{C\alpha}\in\mathbb{R}^{N_\mathrm{res}\times 3}$ denotes the collection of alpha-carbon coordinates of the protein; %
    \item $\mathbf{x}_\mathrm{nonC\alpha}\in\mathbb{R}^{(N_\mathrm{protatm}-N_\mathrm{res})\times 3}$ denotes the set of coordinates for all non-alpha-carbon protein atoms (backbone N, C, O, and all side-chain heavy atoms);
    \item $\mathbf{y}\in \mathbb{R}^{N_\mathrm{ligatm}\times 3}$ denotes all ligand heavy atom coordinates. Note that $N_\mathrm{ligatm}\defeq\sum_{k=1}^K N_{\mathrm{ligatm}, k}$ with $N_{\mathrm{ligatm}, k}$ being the number of heavy atoms in each ligand molecule ${G}_k$.
\end{itemize}
$\mathbf{U}$ defines a linear transformation such that %
the forward diffusion process erases chemical-group-scale local details before it removes global information about protein domain packing and ligand binding interfaces. $\mathbf{U}$ is defined through
\begin{equation}
    \label{eq:U_transform}
    \mathbf{z}= \mathbf{U}^{-1} \mathbf{Z} = \big[ \mathbf{x}_\mathrm{C\alpha},\ \mathbf{x}_\mathrm{nonC\alpha} - \mathbf{x}_\mathrm{C\alpha},\ \mathbf{y} - \mathbf{c}^\top \mathbf{x}_\mathrm{C\alpha} \big]^\top 
\end{equation}
where $\mathbf{c}$ is the softmax-transformed contact map $c_{A, M}=\frac{\sum_{J\in\{M\}}\exp(L_{AJ})}{\sum_{A, J\in\{M\}} \exp(L_{AJ})}$, where $M\in\{1, \cdots, N_\mathrm{ligands}\}$ is a ligand index running over all molecular graphs in $\{G\}$; $\bm{\lambda}$ is a diagonal matrix representing the drift coefficients on latent coordinates $\mathbf{z}$. The diagonal entries of $\bm{\lambda}$ are set to 6.0 for alpha-carbon atoms $\mathbf{x}_\mathrm{C\alpha}$ and 37.5 for all other internal coordinates $\mathbf{x}_\mathrm{nonC\alpha} - \mathbf{x}_\mathrm{C\alpha},\ \mathbf{y} - \mathbf{c}^\top \mathbf{x}_\mathrm{nonC\alpha}$. 
We set the diffusion coefficient parameter $\sigma=\SI{12.25}{\angstrom}$.
For all SDE sampling, we use an exponential time schedule defined by
\begin{equation}
    \label{eq:DiffusionTimeSchedule}
    t = e^{-\log(t_0) \cdot \tau} \cdot t_0 \quad \textrm{where} \quad  t_0 = 10^{-3},\ \tau \sim (0, 1.0 ]
\end{equation}
analogous to common choices in variance-exploding~\cite{song2020score} SDEs.
The marginal densities of the latent coordinates $\mathbf{z}$ sampled from the forward SDE are given by: 
\begin{equation}
    \label{eq:forward_kernel}
    q(\mathbf{z}_t | \mathbf{z}_0) = \mathcal{N}(\sqrt{\bm{\alpha}_t} \cdot \mathbf{z}_0,\ (\mathbf{I} - \bm{\alpha}_t) \cdot \frac{\sigma^2}{2 \bm{\lambda}}) \quad \textrm{where} \quad \bm{\alpha}_t = \exp(-2 \bm{\lambda} t)
\end{equation}
Proofs regarding equivariance and chiral properties are stated in the Supplemental Information,~\ref{sec:sde_se3}.

\subsection*{Temperature-annealed and continuity-corrected diffusion model sampling}

\paragraph{Langevein simulated annealing SDE (LSA-SDE)}

\label{sec:LSA}

To enable the generation of crystal-structure-like complexes corresponding to local minima of the model density distribution, we adopt an adjusted diffusion model sampling scheme where an auxiliary inverse temperature parameter $\beta(t)$ is smoothly varied from 1.0 (i.e., the data distribution temperature) to a target inverse temperature $\beta_\mathrm{max}=10.0$ along the reverse-time SDE. Inspired by recent works on controllable diffusion-based generation for protein backbones~\cite{Ingraham2022.12.01.518682}, we introduce the following dynamics termed Langevin simulated annealing SDE (LSA-SDE):
\begin{equation}
    \label{eq:lsa_sde}
    d\mathbf{Z}_t = [-\Theta \mathbf{Z}_t -\frac{1+\beta(t)}{2} \Sigma \Sigma^\top \nabla_{\mathbf{Z}_t} \log{q_t(\mathbf{Z}_t)}] dt + \sqrt{\frac{1}{\beta(t)}} \Sigma_t d\mathbf{W}_t
\end{equation}
where we linearly varied the inverse temperature parameter $\beta(t)$ according to $\beta(t) = 1 + (\beta_\mathrm{max} - 1) (1 - \tau)$.  

\paragraph{Kabsch alignment correction} {While marginal densities from the forward SDE depend on the global placement and orientation of the initial molecular structure, neural-network-based structure prediction is mainly found to achieve improved accuracy when trained with SE(3)-invariant losses~\cite{jumper2021highly}. However, structure prediction models trained with invariant loss can cause inconsistency between integration steps when used for diffusion model sampling due to the arbitrariness of predicted global rotation, which is found to harm the generation quality in related works~\cite{rfdiffusion}. To reconcile this discrepancy, at sampling time we introduce a rigid-body alignment operation between the C$\alpha$ traces of the current and previous-step network outputs using the Kabsch-Umeyama algorithm and then use the aligned structure to compute the score function $\mathbf{s}_{\phi}$: 
\begin{equation}
    \mathbf{s}_{\phi}(\mathbf{Z}_t, t) = \mathbf{U} \frac{1}{1 - \bm{\alpha}_t} \big(\sqrt{\bm{\alpha}_t} \cdot \mathbf{U}^{-1}  (\mathbf{R} \circ \hat{\mathbf{Z}}_0) -  \mathbf{U}^{-1} \mathbf{Z}_t \big), \quad \mathbf{R} = \mathrm{argmin}_{\mathbf{r}\in\mathrm{SE(3)}} \lVert \mathbf{r} \circ \hat{\mathbf{x}}_{\mathrm{C\alpha},0}(\mathbf{Z}_t) - \hat{\mathbf{x}}_{\mathrm{C\alpha},0}(\mathbf{Z}_{t+\Delta t}) \rVert^2_2
\end{equation}
where we have assumed a time discretization interval of $\Delta t$ for a single integrator step.

\paragraph{Semi-analytic integrator}

Furthermore, we introduce a semi-analytic integration scheme based on a harmonic approximation of the time-dependence score function $\mathbf{s}_{\phi}$ to reduce discretization errors:
\begin{equation}
    \label{eq:sa_step}
    \mathbf{z}_t = \sqrt{\bm{\alpha}_t} \Big[1 - \big(\frac{1/\bm{\alpha}_t - 1}{1/\bm{\alpha}_{t+\Delta t} - 1}\big)^{\frac{1+\beta(t)}{4 \bm{\lambda}}\sigma^2} \Big] \mathbf{R} \circ \hat{\mathbf{z}}_0(\mathbf{Z}_{t+\Delta t}) + \big(\frac{1/\bm{\alpha}_t - 1}{1/\bm{\alpha}_{t+\Delta t} - 1}\big)^{\frac{1+\beta(t)}{4 \bm{\lambda}}\sigma^2} \sqrt{\frac{\bm{\alpha}_t}{\bm{\alpha}_{t+\Delta t}}} \mathbf{z}_{t+\Delta t}  + \sigma \sqrt{\frac{\bm{\alpha}_{t} - \bm{\alpha}_{t+\Delta t}}{ 2 \bm{\lambda} \beta(t)}} \bm{\epsilon}
\end{equation}
where $\bm{\epsilon}\sim \mathcal{N}(0,\ \mathbf{I})$.
Equation \eqref{eq:sa_step} can be derived by directly applying Ito's Lemma to a linearized form of the reverse-time SDE, as discussed in detail in Supplemental Information, Section~\ref{sec:sde}. 

\subsection*{Datasets and training summary} The datasets used for training and testing end-to-end structure prediction were constructed from chains of all monomeric proteins and homomeric complexes in the Protein Data Bank accessed on April 2022. We filtered the retrieved structures by discarding structures with experimental resolution lower than \SI{3.0}{\angstrom} and chain coverage lower than 95\%, and deleting ligands that contain more than 1000 heavy atoms and non-transition-metal single-atom ligands. The holo structures are obtained from the intersection between the filtered set and the annotated protein-ligand complex database BioLip~\cite{yang2012biolip} using their non-redundant index file as well as an extra set of chains from GPCRdb~\cite{pandy2023gpcrdb}, and dropping samples with DNA and RNA ligands; the apo structures are obtained from the filtered set that contains no ligand or only ligands in the artifact list provided by Ref.~\cite{yang2012biolip}. After combining the apo and holo structures and removing duplicate chains, we obtained 85140 unique samples. The final PL2019-74k dataset is obtained by removing samples deposited after Jan 2019 and samples with UniProt ID in the PocketMiner dataset~\cite{meller2022predicting}, resulting in 74477 samples for model training. The contrasting apo-holo pair test set is then obtained by taking the samples of the same PDB code and chain index in the original PocketMiner dataset; the recent receptors test set is obtained by taking the filtered samples deposited in Jan 2019 for which a reference experimental sample with sequence identity > 98\%, %
backbone RMSD>\SI{2}{\angstrom} as reported by TMAlign~\cite{zhang2005tm}, and distinct bound ligands can be found from the PDB. 

During training, we perform a weighted subsampling on the PL2019-74k dataset at each epoch to balance the occurrence frequency of unique UniRef50 cluster IDs and maximum backbone RMSD to PDB structures. With a 50\% probability at each training iteration, an experimental template structure or an AlphaFold2-predicted structure of pLDDT>0.7 or higher is supplied to NeuralPLexer inputs as an auxiliary template structure. Those AlphaFold2 structures were obtained with ColabFold~\cite{mirdita2022colabfold}; for each unique chain in PL2019-74k, we sample at most 20 structures using input MSAs provided by OpenFold~\cite{ahdritz2022openfold} with reduced MSA sizes of [16, 32, 64, 128, 256] according to the bootstrapping technique described in Ref.~\cite{del2022sampling} and without structure template inputs. 
Additional details related to loss functions, the training protocol, and hyperparameters can be found in Supplementary Information, Section~\ref{sec:si_training}.

\subsection*{Weighted Q factor} %
The weighted Q factor metric between a predicted protein structure $\mathbf{x}_\mathrm{pred}$ and a reference protein structure $\mathbf{x}_\mathrm{alt}$ is defined based on comparing the overlap between the two contact maps, over residue pairs for which the contacts are not conserved between the reference structure and an alternative conformation of the protein $\mathbf{x}_\mathrm{alt}$:
\begin{equation}
    \mathrm{WeightedQFactor}(\mathbf{x}_\mathrm{pred}, \mathbf{x}_\mathrm{ref}; \mathbf{x}_\mathrm{alt}) = \frac{\sum_{A,B} (\mathbf{C}_{\mathrm{pred},AB} \land \mathbf{C}_{\mathrm{ref},AB}) \cdot [1 - (\mathbf{C}_{\mathrm{ref},AB} \land \mathbf{C}_{\mathrm{alt},AB})] \cdot  \mathbbm{1}_{ | d_{\mathrm{ref},AB} - d_{\mathrm{alt},AB} | > \SI{2.0}{\angstrom}} }{\sum_{A,B} [1 - (\mathbf{C}_{\mathrm{ref},AB} \land \mathbf{C}_{\mathrm{alt},AB})]  \cdot  \mathbbm{1}_{ | d_{\mathrm{ref},AB} - d_{\mathrm{alt},AB} | > \SI{2.0}{\angstrom}}}
\end{equation}
where $\mathbbm{1}$ denotes a 0-1 indicator function, and $\land$ denotes a logical and operation; the contact map $\mathbf{C}$ is defined as a binary matrix: for two residues $A, B$, the matrix element $\mathbf{C}_{AB}=1$ if the distance $d_{AB}$ between the centroid coordinates of residue $A$ and $B$ is closer than $\SI{20}{\angstrom}$, and $\mathbf{C}_{AB}=0$ otherwise. For all results reported in Figures~\ref{fig:fig4}-\ref{fig:fig5}, the alternative structure $\mathbf{x}_\mathrm{alt}$ is set to the most contrasting sample from the PL2019-74k dataset, that is, the structure with highest backbone RMSD as reported by TMAlign~\cite{zhang2005tm} among samples with sequence identity > 98\% to the query structure.

\section*{Acknowledgements}
Z.Q. acknowledges graduate research funding from Caltech and partial support from the Amazon–Caltech AI4Science fellowship. T.M. acknowledges partial support from the Caltech DeLogi fund, and A.A. acknowledges support from a Caltech Bren professorship. We thank Matthew Welborn, Chao Zhang, and Vignesh Bhethanabotla for helpful discussions on the work and for comments on the manuscript. We thank Artur Meller and Jonathan Borowsky for sharing the PocketMiner dataset.

\printbibliography

\begin{refsection}

\newpage
\setcounter{figure}{0}
\renewcommand{\thefigure}{S\arabic{figure}}
\setcounter{table}{0}
\renewcommand{\thetable}{S\arabic{table}}
\setcounter{algorithm}{0}
\renewcommand{\thealgorithm}{S\arabic{algorithm}}
\setcounter{equation}{0}
\renewcommand{\theequation}{S\arabic{equation}}
\newrefcontext[labelprefix=S]

\appendix

\begin{center}
    \textbf{\huge Supplementary Information}
\end{center}

\tableofcontents

\renewcommand{\algorithmiccomment}[1]{\hfill #1}

\section{Algorithm overview}

\begin{algorithm}
\caption{NeuralPLexer Inference}
\label{alg:main}
\begin{algorithmic}[1]
\REQUIRE $\{\mathbf{s}\}$, $\{{G}\}$, $N_\mathrm{conformations}$, $N_\mathrm{steps}=40$, $use\_template$, $compute\_plddt$
\STATE $\{\mathbf{f}_\mathrm{PLM} \} \gets$ Compute ESM-2-650M features for all input chains in $\{\mathbf{s}\}$ \COMMENT{Ref.~\cite{lin2023evolutionary}}
\IF{$use\_template$}
    \STATE $\{ \mathbf{f}_\mathrm{template} \} \gets$ Retrieve template protein structure and compute template features 
\ENDIF
\FOR{$i \in \{1, \cdots, N_\mathrm{conformations}\}$}
\STATE $T = \mathrm{DiffusionTimeSchedule(\tau=1.0)}$ \COMMENT{Equation~\eqref{eq:DiffusionTimeSchedule}}
\STATE Sample initial protein coordinates $\mathbf{x}_{T}$ from the prior $q_{T}$ \COMMENT{Equation~\eqref{eq:prior_sampling}}
\STATE $residue\_graph\_0 \gets$ Generate the residue-scale graph based on $(\mathbf{x}_T)_\mathrm{C\alpha}$ \COMMENT{Section~\ref{sec:graph_construction}}
\STATE $\mathbf{F}_\mathrm{P} \gets$ Sample $N_\mathrm{P}$ protein backbone frame nodes and gather embeddings  \COMMENT{$N_\mathrm{P}=96$}
\STATE $\tilde{\mathbf{l}} \gets \mathbf{0}$
\\ \textit{\# Generating contact maps and block-adjacency matrices (Equation~\eqref{eq:contact_forward})} 
\IF{$\{{G}\}$ contains more than 0 ligands}
    \STATE Compute MHT embeddings for all input ligand molecular graphs in $\{{G}\}$ \COMMENT{Algorithm~\ref{alg:MHT}}
    \STATE  $\mathbf{F}_\mathrm{L} \gets$ Sample $N_\mathrm{L}$ ligand frame nodes, gather and symmetrize MHT embeddings \COMMENT{$N_\mathrm{L}=32$}
    \FOR{$k \in \{1, \cdots, N_\mathrm{L}\}$}
        \STATE $residue\_graph\_k \gets \mathrm{CPMForward}(residue\_graph\_0, \tilde{\mathbf{l}}, \tau=1.0)$ \COMMENT{Section~\ref{sec:CPM}}
        \STATE $\hat{P} = \mathrm{LinearNoBias}(\mathrm{MLP}(\mathrm{GetEdges_{BF}}(residue\_graph\_k)))$ \COMMENT{$\hat{P} \in \mathbb{R}^{N_\mathrm{res}\times N_\mathrm{L} \times N_\mathrm{bins}}, N_\mathrm{bins}=32$}
        \STATE $\tilde{L}_k \gets \mathrm{DistogramToContactMap}(\hat{P})$ \COMMENT{Equation~\eqref{eq:DistogramToContactMap}}
        \STATE $\hat{\mathbf{L}}_k \gets \mathrm{SumIntoPatches}(\tilde{L}_k) \odot [1 - \mathrm{max}_\mathrm{column-wise}(\tilde{\mathbf{l}})]$ \COMMENT{$\hat{\mathbf{L}}_k \in \mathbb{R}^{N_\mathrm{P}\times N_\mathrm{L} }$}
        \STATE $\mathbf{l}_k = \mathrm{OneHot}(i_k, j_k); \ (i_k, j_k)\sim\mathrm{Categorical}_{N_\mathrm{P}\times N_\mathrm{L}}(\hat{\mathbf{L}}_{k})$  \COMMENT{$\mathbf{l}_k \in \{0, 1\}^{N_\mathrm{P}\times N_\mathrm{L} }$}
        \STATE $\tilde{\mathbf{l}} \gets \tilde{\mathbf{l}} + \mathbf{l}_k$  \COMMENT{$\tilde{\mathbf{l}} \in \{0, 1\}^{N_\mathrm{P}\times N_\mathrm{L} }$}
    \ENDFOR
    \STATE Compute $\mathbf{c}$ and $\mathbf{U}$ for ligand degrees of freedom using predicted $\hat{\mathbf{L}}\defeq \hat{\mathbf{L}}_{N_\mathrm{L}}$  \COMMENT{Equation~\eqref{eq:U_transform}}
    \STATE Sample initial ligand coordinates $\mathbf{y}_T$ from the prior $q_{T}$ \COMMENT{Equation~\eqref{eq:prior_sampling}}
    \STATE $\mathbf{Z}_T \gets \mathrm{concat}(\mathbf{x}_T, \mathbf{y}_T)$
    \ELSE
    \STATE $\mathbf{Z}_T \gets \mathbf{x}_T$
\ENDIF
\\ \textit{\# Generating the 3D structure for all heavy atoms (Equation~\eqref{eq:reverse_sde})}
\STATE Compute MHT embeddings for all amino acid molecular graphs \COMMENT{Algorithm~\ref{alg:MHT}}
\FOR[$\Delta \tau = 1/N_\mathrm{steps}$]{$\tau \in \{1, 1-\Delta \tau, \cdots, \Delta \tau \}$}
    \STATE $t \gets \mathrm{DiffusionTimeSchedule(\tau)}$ \COMMENT{Equation~\eqref{eq:DiffusionTimeSchedule}}
    \STATE $\Delta t \gets \mathrm{DiffusionTimeSchedule(\tau)} - \mathrm{DiffusionTimeSchedule(\tau-\Delta \tau)}$ \COMMENT{Equation~\eqref{eq:DiffusionTimeSchedule}}
    \STATE $residue\_graph, atomic\_graph \gets$ Regenerate the residue-scale and atomic-scale graph based on $\mathbf{Z}_t$ \COMMENT{Section~\ref{sec:graph_construction}}
    \STATE $residue\_graph \gets \mathrm{CPMForward}(residue\_graph, \tilde{\mathbf{l}}, \tau)$ \COMMENT{Section~\ref{sec:CPM}}
    \STATE $graph\_rep \gets \mathrm{Collate}(residue\_graph, atomic\_graph, cross\_scale\_graph)$ \COMMENT{Table~\ref{table:all_edges}}
    \STATE $\hat{\mathbf{Z}}_0 \gets \mathrm{ESDMForward}(\mathbf{Z}_t; graph\_rep, \tau)$ \COMMENT{Section~\ref{sec:ESDM}}
    \STATE $\beta_{t-\Delta t} \gets \mathrm{InvTempSchedule}(\tau-\Delta \tau)$ \COMMENT{See Methods, LSA-SDE}
    \STATE $\mathbf{Z}_{t-\Delta t} = \mathrm{IntegratorStep}(\mathbf{Z}_t, \hat{\mathbf{Z}}_0, \beta_{t-\Delta t}, \Delta t)$ \COMMENT{Equation~\eqref{eq:sa_step}}
\ENDFOR
\\ \textit{\# Assigning confidence estimations (per-residue and per-ligand-atom pLDDT)}
\IF{$compute\_plddt$}
    \STATE $residue\_graph, atomic\_graph \gets$ Regenerate the residue-scale and atomic-scale graph based on $\mathbf{Z}_t$ \COMMENT{Section~\ref{sec:graph_construction}}
    \STATE $residue\_graph \gets \mathrm{CPMForward}(residue\_graph, \mathbf{0}, \tau=0.0)$ \COMMENT{Section~\ref{sec:CPM}}
    \STATE $graph\_rep \gets \mathrm{Collate}(residue\_graph, atomic\_graph, cross\_scale\_graph)$ \COMMENT{Table~\ref{table:all_edges}}
    \STATE $pLDD\_gram = \mathrm{ConfidenceEstimationHead}(\mathbf{Z}_0; graph\_rep)$ \COMMENT{Section~\ref{sec:si_plddt}}
    \STATE $pLDDT\_prot, pLDDT\_lig = \mathrm{compute\_plDDT}(pLDD\_gram)$ \COMMENT{Equation~\ref{eq:plddt}}
    \STATE \textbf{yield} {$\mathbf{Z}_0, (pLDDT\_gram, pLDDT\_prot, pLDDT\_lig)$}
    \ELSE 
    \STATE \textbf{yield} $\mathbf{Z}_0$
\ENDIF
\ENDFOR
\end{algorithmic}
\end{algorithm}

Given the set of query protein sequences, ligand molecular graphs, and optional template protein structure inputs, Algorithm~\ref{alg:main} summarizes the process that NeuralPLexer uses to sample an ensemble of protein or protein-ligand complex structures.

Unless stated otherwise, $\mathrm{LinearNoBias(\cdot)}$ denotes a standard linear transformation with
a trainable weight matrix right-multiplied to the last input tensor dimension; $\mathrm{MLP}(\cdot)$ denotes a standard 3-layer multilayer perceptron with GELU activation function~\cite{hendrycks2016gaussian} and with layer normalization~\cite{ba2016layer} applied to the output activations.

\section{Diffusion SDEs}

\label{sec:sde}

\subsection{Preliminaries}

For the temporally homogeneous forward-time SDE
\begin{equation}
\label{eq:frdsde}
    d\mathbf{Z}_t = -(\mathbf{U} \bm{\lambda} \mathbf{U}^{-1} ) \mathbf{Z}_t dt + (\sigma \mathbf{U}) d\mathbf{W}_t
\end{equation}
or equivalently, in the space of latent coordinates
\begin{equation}
\label{eq:frdsde_lat}
    d\mathbf{z}_t = d(\mathbf{U}^{-1} \mathbf{Z}_t) = -\bm{\lambda} \cdot \mathbf{z}_t dt + \sigma  d\mathbf{W}_t
\end{equation}
the corresponding reverse-time SDE is given by
\begin{equation}
\label{eq:revsde}
    d\mathbf{Z}_t = [-(\mathbf{U} \bm{\lambda} \mathbf{U}^{-1} ) \mathbf{Z}_t  - \sigma^2 \mathbf{U}\mathbf{U}^\top \mathbf{s}_{\phi}(\mathbf{Z}_t; t)]dt + (\sigma \mathbf{U}) d\mathbf{W}_t \ .
\end{equation}
The initial coordinates at $t=T=1.0$ are sampled from the prior distribution $q_T$
\begin{equation}
    \mathbf{Z}_T \sim q_T \defeq \mathcal{N}(\mathbf{0},\ \mathbf{U} \frac{\sigma^2}{2 \bm{\lambda}} \mathbf{U}^{\mathrm{T}})
\end{equation}
or equivalently,
\begin{equation}
    \label{eq:prior_sampling}
    \mathbf{Z}_T = \mathbf{U} \sqrt{\sigma/2} \, \bm{\lambda}^{-\frac{1}{2}} \bm{\varepsilon}, \quad \bm{\varepsilon} \sim \mathcal{N}(\mathbf{0},\ \mathbf{I})
\end{equation}

\subsection{Euclidean and chiral symmetries}
\label{sec:sde_se3}

Given group $G$, a function $f: X \rightarrow Y$ is said to be equivariant if for all $g\in G$ and $x \in X$, $f(\varphi_X(g) \cdot x) = \varphi_Y(g) \cdot f(x)$, and $f$ is said to be invariant if $f(\varphi_X(g) \cdot x) = f(x)$. We are interested in the special Euclidean group $G=\mathrm{SE(3)}$ which consists of all global rigid translation and rotation operations $g\cdot \mathbf{Z} \defeq \mathbf{t} + \mathbf{Z} \cdot \mathbf{R}$ where $\mathbf{t}\in\mathbb{R}^3$ and $\mathbf{R} \in \mathrm{SO(3)}$. To adhere to the physical constraint that $p_\mathrm{data}$ is always $\mathrm{SE(3)}$-invariant, %
the transition kernels of the forward-time SDE need to satisfy $\mathrm{SE(3)}$-equivariance $q(\mathbf{Z}_{t+s} | \mathbf{Z}_t ) = q( \varphi_{t+s}(g) \cdot \mathbf{Z}_{t+s} | \varphi_{t}(g) \cdot \mathbf{Z}_t )$ such that all marginals are invariant $q_t(\mathbf{Z}_t) = q_t(\varphi_{t}(g) \cdot \mathbf{Z}_t)$ for all diffusion times $t$. Since the forward SDE only involves terms that are isotropic in (x, y, z) components, the proof is straightforward:
\begin{align*}
    &q(\sqrt{\bm{\alpha}_s} \mathbf{t} + \mathbf{Z}_{t+s} \cdot \mathbf{R} \vert \mathbf{t} + \mathbf{Z}_{t} \cdot \mathbf{R})\\
    &= \mathcal{N}\big(\sqrt{\bm{\alpha}_s} \mathbf{t} + \mathbf{Z}_{t+s} \cdot \mathbf{R};\ \sqrt{\bm{\alpha}_s} \big( \mathbf{t} + \mathbf{Z}_{t} \cdot \mathbf{R} \big), \mathbf{U} \frac{\sqrt{\mathbf{I} - \bm{\alpha}_s} \sigma^2}{2 \bm{\lambda}} \mathbf{U}^{\mathrm{T}}  \big)\\ 
    &= \mathcal{N}\big(\sqrt{\bm{\alpha}_s} \mathbf{t} \mathbf{R}^\top + \mathbf{Z}_{t+s} \cdot \mathbf{R} \mathbf{R}^\top;\ \sqrt{\bm{\alpha}_s} \big( \mathbf{t} \mathbf{R}^\top + \mathbf{Z}_{t} \cdot \mathbf{R} \mathbf{R}^\top \big), \mathbf{U} \frac{\sqrt{\mathbf{I} - \bm{\alpha}_s} \sigma^2}{2 \bm{\lambda}} \mathbf{U}^{\mathrm{T}} \otimes \mathbf{R} \mathbf{R}^\top \big)\\ 
    &= \mathcal{N}\big(\mathbf{Z}_{t+s};\ \mathbf{Z}_{t}, \mathbf{U} \frac{\sqrt{\mathbf{I} - \bm{\alpha}_s} \sigma^2}{2 \bm{\lambda}} \mathbf{U}^{\mathrm{T}} \big)\\ 
    &= q(\mathbf{Z}_{t+s} \vert \mathbf{Z}_{t} ) .
\end{align*}
where the translation term after a time interval $s$ is scaled by a factor $\sqrt{\bm{\alpha}_s}$ due to the linear drift term in Eq.~\eqref{eq:frdsde}.
Because all transition kernels are $\mathrm{SE(3)}$-equivariant, it then follows that the score function $\nabla_{\mathbf{Z}} \log q_t(\mathbf{Z})$ is also equivariant: 
\begin{align*}
    & \nabla_{\mathbf{Z}'} \log q_t(\mathbf{Z}') \quad \mathrm{where} \quad \mathbf{Z}'= \sqrt{\bm{\alpha}_t}\mathbf{t}+\mathbf{Z}\cdot \mathbf{R} \\
    & = \nabla_{\mathbf{Z}'} \log [\mathbb{E}_{\mathbf{Z}_0 \sim q_\mathrm{data}} q_t(\mathbf{Z}' \vert \mathbf{Z}_0 ) ] \\
    & = \nabla_{\mathbf{Z}'} \log [\mathbb{E}_{\mathbf{Z}_0 \sim q_\mathrm{data}} q_t(\mathbf{Z}' \vert \mathbf{t}+\mathbf{Z}_0 \cdot \mathbf{R} ) ] \\
    & = \nabla_{\mathbf{Z}'} \log [\mathbb{E}_{\mathbf{Z}_0 \sim q_\mathrm{data}} q_t(\mathbf{Z} \vert \mathbf{Z}_0 ) ] \\
    & = \frac{\partial \mathbf{Z}}{\partial \mathbf{Z}'} \nabla_{\mathbf{Z}} \log [\mathbb{E}_{\mathbf{Z}_0 \sim q_\mathrm{data}} q_t(\mathbf{Z} \vert \mathbf{Z}_0 ) ] \\
    &= \nabla_{\mathbf{Z}} \log q_t(\mathbf{Z}) \cdot \mathbf{R}.
\end{align*}

While the forward SDE is $\mathrm{E(3)}$-equivariant as the noising process satisfies $q(-\mathbf{Z}(t + s) | -\mathbf{Z}(t)) = q(\mathbf{Z}(t + s) | \mathbf{Z}(t))$, it is worth noting that the reverse SDE is only $\mathrm{SE(3)}$-equivariant as parity-inversion transformations $i: \mathbf{Z} \mapsto - \mathbf{Z}$ on the data distribution $p_\mathrm{data}$ is physically forbidden and thus the score $\nabla_{\mathbf{Z}} \log q_t(\mathbf{Z})$ is of broken chiral symmetry; in general: 
there exists $\mathbf{Z}$ such that $\nabla_{\mathbf{-Z}} \log q_t(\mathbf{-Z}) \neq -\nabla_{\mathbf{Z}} \log q_t(\mathbf{Z})$.

\subsection{Derivation of the semi-analytic integrator (Eq.~\ref{eq:sa_step})}

Without loss of generality, we consider a homogeneous, fixed-temperature analog of the temperature-adjusted reverse-time SDE~\eqref{eq:lsa_sde} with $\mathbf{z}_t \in \mathbb{R}^3$, $t\in (0, \infty )$
\begin{equation}
    d\mathbf{z}_t = [- \theta \mathbf{z}_t - \frac{1+\beta}{2} \sigma^2  \mathbf{s}_\phi(\mathbf{z}_t; t)] dt + \sqrt{\frac{1}{\beta}} \sigma  d\mathbf{W}_t
\end{equation}
where
\begin{equation}
    \mathbf{s}_\phi(\mathbf{z}_t; t) = \frac{\sqrt{\alpha_t} \hat{\mathbf{z}}_0(\mathbf{z}_t) - \mathbf{z}_t}{1 - \alpha_t} ; \quad \alpha_t = \exp(-2\theta t)
\end{equation}
thus 
\begin{equation}
     d\mathbf{z}_t = [- \theta \mathbf{z}_t -  \frac{1+\beta}{2} \sigma^2 \frac{\exp(-\theta t) \hat{\mathbf{z}}_0(\mathbf{z}_t) - \mathbf{z}_t}{1 - \exp(-2\theta t)}] dt +  \sqrt{\frac{1}{\beta}}  \sigma  d\mathbf{W}_t
\end{equation}

For an integration time interval from $s=t+\Delta t$ to $t$, we make the approximation that the score function is linear with respect to the attraction term $\hat{\mathbf{z}}_0(\mathbf{z}_s; \phi, s) - \frac{1}{\sqrt{\alpha_{\tau}}} \mathbf{z}_{\tau}$ for $\tau \in [t, s)$:
\begin{equation}
     d \mathbf{z}_{\tau} = [- \theta \mathbf{z}_\tau - \frac{1+\beta}{2}  \sigma^2 \frac{\exp(-\theta \tau) \hat{\mathbf{z}}_0(\mathbf{z}_s) - \mathbf{z}_{\tau}}{1 - \exp(-2\theta {\tau})}] d\tau +  \sqrt{\frac{1}{\beta}}  \sigma  d\mathbf{W}_\tau
\end{equation}
or equivalently,
\begin{equation}
     d \mathbf{z}_{\tau} = a(\tau) + b(\tau) \mathbf{z}_\tau \, d\tau + c \, d\mathbf{W}_\tau
\end{equation}
where 
\begin{equation}
     a(\tau) =  - \frac{1+\beta}{2}  \sigma^2 \frac{\exp(-\theta \tau) \hat{\mathbf{z}}_0(\mathbf{z}_s)}{1 - \exp(-2\theta {\tau})}, \quad b(\tau) = [ \frac{ \frac{1+\beta}{2}  \sigma^2}{1 - \exp(-2\theta {\tau})} - \theta] ; \quad c = \sqrt{\frac{1}{\beta}}  \sigma  .
\end{equation}

Defining $\Psi(t, s)\defeq \exp(\int_s^t -b(\tau) d\tau)$, by applying Ito's Lemma we note that
\begin{align}
    d (\Psi(t, s) \mathbf{z}_t) &= [\frac{d}{d t} \Psi(t, s)  \mathbf{z}_t + a(t) \Psi(t, s) + b(t)  \mathbf{z}_t  \Psi(t, s) ] dt + c \Psi(t, s) d\mathbf{W}_t \\
    &= a(t) \Psi(t, s) dt +  c \Psi(t, s) d\mathbf{W}_t
\end{align}
which implies
\begin{equation}
    \mathbb{E}(\mathbf{z}_t) = \int_s^t a(\tau) \frac{\Psi(\tau, s)}{\Psi(t, s)} d\tau + \frac{\mathbf{z}_s}{\Psi(t, s)}; \quad \mathrm{Var}(\mathbf{z}_t) = \int_s^t \Big( c \frac{\Psi(\tau, s)}{\Psi(t, s)} \Big)^2 d\tau
\end{equation}
Computing the integrals analytically yields
\begin{equation}
    \Psi(t, s) = \Big( \frac{e^{2\theta s} -1 }{e^{2\theta t}-1} \Big)^{\frac{1+\beta}{4\theta} \sigma^2} \cdot e^{\theta (t-s)}
\end{equation}
\begin{align}
    \int_s^t a(\tau) \frac{\Psi(\tau, s)}{\Psi(t, s)} d\tau &= \int_s^t  - \frac{1+\beta}{2}  \sigma^2 \frac{\exp(-\theta \tau) \hat{\mathbf{z}}_0(\mathbf{z}_s)}{1 - \exp(-2\theta {\tau})}  \Big( \frac{e^{2\theta t} -1 }{e^{2\theta \tau}-1} \Big)^{\frac{1+\beta}{4\theta} \sigma^2} \cdot e^{\theta (\tau-t)} d\tau \\
    &= \int_s^t - \frac{1+\beta}{2}  \sigma^2 \frac{ \hat{\mathbf{z}}_0(\mathbf{z}_s)}{1 - \exp(-2\theta {\tau})}  \Big( \frac{e^{2\theta t} -1 }{e^{2\theta \tau}-1} \Big)^{\frac{1+\beta}{4\theta} \sigma^2} \cdot e^{- \theta t} d\tau \\
    &=  - \frac{1+\beta}{2} \sigma^2 \hat{\mathbf{z}}_0(\mathbf{z}_s) e^{- \theta t}  \Big( e^{2\theta t} -1 \Big)^{\frac{1+\beta}{4\theta} \sigma^2}  \int_s^t \frac{e^{2\theta \tau} }{(e^{2\theta \tau}-1)^{1+\frac{1+\beta}{4\theta} \sigma^2}} d\tau \\
    &=  e^{- \theta t}  \Big[ 1  - \Big( \frac{e^{2\theta t} -1 }{e^{2\theta s}-1} \Big)^{\frac{1+\beta}{4\theta} \sigma^2} \Big] \hat{\mathbf{z}}_0(\mathbf{z}_s)
\end{align}
\begin{align}
    \mathbb{E}(\mathbf{z}_t) &=  e^{- \theta t}  \Big[ 1  - \Big( \frac{e^{2\theta t} -1 }{e^{2\theta s}-1} \Big)^{\frac{1+\beta}{4\theta} \sigma^2} \Big] \hat{\mathbf{z}}_0(\mathbf{z}_s) + \Big( \frac{e^{2\theta t} -1 }{e^{2\theta s}-1} \Big)^{\frac{1+\beta}{4\theta} \sigma^2} \cdot e^{\theta (s-t)} \mathbf{z}_s 
\end{align}
For simplicity, we adapt the variance term to that of the forward SDE: $\mathrm{Var}(\mathbf{z}_t) \approx \frac{\sigma^2}{2 \theta \beta} (e^{-2\theta t} - e^{-2\theta s})$. 
Matching the conditional expectations and variances to the Gaussian transition kernel $q_{t:t+\Delta t}(\cdot | \mathbf{z}_{t+\Delta t})$ and directly generalizing to the rotation-corrected multivariate setting, we recover Equation~\ref{eq:sa_step}.
Note that the DDIM~\cite{song2020denoising} integrator can be recovered by removing the noise term and setting $\beta\equiv 0, \sigma \equiv 1, \theta\equiv \frac{1}{2}$ which corresponds to the standard variance-preserving (VP)-SDE~\cite{song2020score}.

\clearpage

\section{Molecular graph featurization and encoder details}

\label{sec:chemfeat}

\subsection{Molecular representations}

Given a set of molecular graphs $\{{G}\}$, the MHT network processes the following collection of embeddings:
\begin{itemize}
    \item Atom representations $\mathbf{f}_\mathrm{atom}\in\mathbb{R}^{N_\mathrm{atom}} \times c$. The input atom representation is a concatenation of one-hot encodings of element group index and period index for the given atom, which is embedded by a linear projection layer $\mathbb{R}^{18+7}\rightarrow\mathbb{R}^c$;
    \item Frame representations $\mathbf{f}_\mathrm{frame}\in\mathbb{R}^{N_\mathrm{frame}} \times c$. For a given frame $u$, $(\mathbf{H}_\mathrm{frame})_u$ is initialized by a 2-layer MLP $\mathbb{R}^{4*2+18+7}\rightarrow\mathbb{R}^c$ that embeds the bond type encodings (defined as $\mathrm{[is\_single, is\_double, is\_triple, is\_aromatic]}$) of the "incoming" bond $(i(u), j(u))$, "outgoing" bond $(j(u), k(u))$, and the atom type encoding of the center atom $j(u)$;
    \item Stereochemistry edge encodings $\mathbf{S}\in\mathbb{R}^{N_\mathrm{frame}\times N_\mathrm{frame} \times c_\mathrm{s}}$, as detailed in Table~\ref{table:stereo}. $\mathbf{S}$ is a sparse tensor where an element $\mathbf{S}_{uv}$ is nonzero only if the pair of frames $(u, v)$ is adjacent, i.e., frame $u$ and frame $v$ sharing a common chemical bond; only pairs with non-zero $\mathbf{S}_\mathrm{uv}$ are included as model inputs.
    \item 3-hop edge representations $\mathbf{f}_\mathrm{aa}\in\mathbb{R}^{N_\mathrm{atom}\times N_\mathrm{atom} \times c_\mathrm{p}}$. For each pair of atoms $(i, j)$, the element $(\mathbf{f}_\mathrm{aa})_{ij}$ is initialized by a linear layer $\mathbb{R}^{4+4}\rightarrow\mathbb{R}^{c_\mathrm{p}}$ that embeds the set of binary graph-distance encodings of whether a path of $k$ ($k\in \{0, 1, 2, 3\}$) chemical bonds exists between atom $i$ and $j$, as well as the bond type one-hot encoding in case a chemical bond exists between atom $i$ and $j$; only edges with non-zero $\mathbf{f}_\mathrm{aa}$ are included as model inputs.
    \item Pair representations $\mathbf{f}_\mathrm{fa}\in\mathbb{R}^{N_\mathrm{frame}\times N_\mathrm{atom} \times c_\mathrm{p}}$. %
    For each frame-atom pair $(u, l)$, the element $(\mathbf{f}_\mathrm{fa})_{ul}$ is initialized by a linear layer $\mathbb{R}^{3*4}\rightarrow\mathbb{R}^{c_\mathrm{p}}$ that embeds the concatenation of graph-distance encodings $\{(\mathbf{f}_\mathrm{aa})_{i(u)l}, (\mathbf{f}_\mathrm{aa})_{j(u)l}, (\mathbf{f}_\mathrm{aa})_{k(u)l}\}$.
\end{itemize}
We denote $\mathbf{X}_\mathrm{s}$ as the $N_\mathrm{frame}\times N_\mathrm{frame}$ binary adjacency matrix of edges among frame nodes, and $\mathbf{X}_\mathrm{a}$ as the $N_\mathrm{atom}\times N_\mathrm{atom}$ binary adjacency matrix of 3-hop edges among atomic nodes. We additionally denote $\mathbf{H}_\mathrm{fa}$ as the $N_\mathrm{frame}\times N_\mathrm{atom}$ incidence matrix between atomic nodes and frame nodes, i.e., $(\mathbf{H}_\mathrm{fa})_{u, l} = 1$ if $l\in\{i(u), j(u), k(u)\}$, and otherwise zero.

Elements of the stereochemistry encoding tensor $\mathbf{S}$ are determined based on the relative orientations among neighboring frames of the input molecular graph.
$\mathrm{is\_above\_plane}(u, v)$ is defined as one of the three atoms in frame $v$ is above the plane formed by frame $u$ with normal vector $\mathbf{v}_u = \frac{(\mathbf{r}_{j(u)}-\mathbf{r}_{i(u)})\times (\mathbf{r}_{k(u)}-\mathbf{r}_{j(u)})}{\lVert \mathbf{r}_{j(u)}-\mathbf{r}_{i(u)}\rVert \lVert \mathbf{r}_{k(u)}-\mathbf{r}_{j(u)} \rVert}$; $\mathrm{is\_same\_side}(u, v)$ is true iff the two bonds not shared between $u, v$ are on the same side of the common bond, equivalent to $\mathbf{v}_u \cdot \mathbf{v}_v > 0$, or vice versa. Our current technical implementations for $\mathrm{is\_above\_plane}$ and $\mathrm{is\_same\_side}$ are based on computing the normal vectors and dot-products using the coordinates from an auxiliary conformer, but we note that in principle all  stereochemistry encodings can be generated based on cheminformatic rules without explicitly generating all atomic coordinates.

\begin{table}[H]
\centering
\caption{Stereochemistry encoding definitions.}
\label{table:stereo}
\begin{tabular}{ll}
\toprule
Feature & Definition \\
\midrule
\multicolumn{2}{l}{\# \textit{Relative topological orientation between two frames}} \\ 
$\mathbf{S}_{uv,0}$ & $\mathrm{common\_bond(u, v) = incoming\_bond(u)}$ \\
$\mathbf{S}_{uv,1}$ & $\mathrm{common\_bond(u, v) = incoming\_bond(v)}$ \\
$\mathbf{S}_{uv,2}$ & $\mathrm{common\_bond(u, v) = outgoing\_bond(u)}$ \\
$\mathbf{S}_{uv,3}$ & $\mathrm{common\_bond(u, v) = outgoing\_bond(v)}$ \\
\multicolumn{2}{l}{\# \textit{Detect small ring structures}} \\ 
$\mathbf{S}_{uv,4}$ & $\mathrm{i(v) \in \{i(u), j(u), k(u) \}}$ \\
$\mathbf{S}_{uv,5}$ & $\mathrm{j(v) \in \{i(u), j(u), k(u) \}}$ \\
$\mathbf{S}_{uv,6}$ & $\mathrm{k(v) \in \{i(u), j(u), k(u) \}}$ \\
\multicolumn{2}{l}{\# \textit{Polyhedral chiral center stereochemistry}} \\ 
$\mathbf{S}_{uv,7}$ & $\mathrm{(j(u)=j(v)) \land is\_above\_plane(u, v)}$ \\
$\mathbf{S}_{uv,8}$ & $\mathrm{(j(u)=j(v)) \land is\_below\_plane(u, v)}$ \\
\multicolumn{2}{l}{\# \textit{Planar stereochemistry for double and $\pi$ bonds}} \\ 
$\mathbf{S}_{uv,9}$ & $\mathrm{is\_double\_or\_aromatic(common\_bond(u, v)) \lor is\_same\_side(u, v)}$ \\
$\mathbf{S}_{uv,10}$ & $\mathrm{is\_double\_or\_aromatic(common\_bond(u, v)) \lor not\_same\_side(u, v)}$ \\
\bottomrule
\end{tabular}
\end{table}

The notion of "frames" in a coordinate-free topological molecular graph is justified by the observation that most bending and stretching modes in molecular vibrations are of high frequency, i.e., most bond lengths and bond angles fall into a small range as predicted by valence bond theory, such that the local frames comprise a consistent molecular representation without prior knowledge of 3D coordinates. 

\subsection{The MHT network architecture}

The forward pass of the Molecular Heat Transformer (MHT) propagates both the node and edges of a graph representation. After executing the MHT network for all ligands in $\{G\}$, we proceed by uniformly sampling $N_\mathrm{L}$ anchor nodes from all $N_\mathrm{frame}$ nodes for subsequent processing; we denote $\mathbf{H}_\mathrm{L}$ as the $N_\mathrm{L} \times N_\mathrm{frame}$ incidence matrix indicating whether a frame node is selected.
\begin{algorithm}[H]
\caption{Molecular Heat Transformer (MHT) Inference}
\label{alg:MHT}
\textbf{def} \textsc{MHT}($\mathbf{f}_\mathrm{atom}, \mathbf{f}_\mathrm{frame}, \mathbf{f}_\mathrm{aa}, \mathbf{f}_\mathrm{fa}, \mathbf{S}, N_\mathrm{blocks}=8, K=8$, $c=512$, $c_\mathrm{p}=64$)
\begin{algorithmic}[1]
\FOR{$i = 1$ to $N_\mathrm{blocks}$} 
    \STATE \textit{\# Pair update block (Line 3-7)}
    \STATE $\mathbf{f}_\mathrm{K}, \mathbf{f}_\mathrm{Q}, \mathbf{b} = \mathrm{LinearNoBias}(\mathbf{f}_\mathrm{frame}), \mathrm{LinearNoBias}(\mathbf{f}_\mathrm{frame}), \mathrm{LinearNoBias}(\mathbf{S})$ \\
    \textit{\# Computing a normalized affinity matrix for nearest-neighbor frames}
    \STATE $\mathbf{U}_{l} = \mathrm{Softmax}_\mathrm{row-wise}\big( \frac{1}{\sqrt{c_\mathrm{P}}}(\mathbf{f}_\mathrm{K} \cdot \mathbf{f}_\mathrm{Q}^\top) + \mathbf{b} - \mathrm{Inf} \cdot (1-\mathbf{X}_\mathrm{s}) \big)$ \\
    \textit{\# Propagating to all frame-frame pairs via heat kernel expansion} \\
    \textit{\# Approximating the matrix exponential $\exp(\mathbf{U}) \approx (\mathbf{1} + \frac{1}{K} \mathbf{U})^K$}
    \STATE $\tilde{\mathbf{U}} = (\mathbf{I} + \frac{1}{K} \mathbf{U}_{l})^K $  \\
    \textit{\# Updating frame-atom pairs using broadcasted kernel matrices}
    \STATE $\mathbf{g} = \tilde{\mathbf{U}} \cdot \mathrm{LinearNoBias}(\mathbf{f}_\mathrm{fa})$ \\
    \STATE $\mathbf{f}_\mathrm{fa} \gets \mathrm{MLP}(\mathrm{concat}(\mathbf{g}, \mathbf{f}_\mathrm{fa})) + \mathbf{f}_\mathrm{fa}$ \\
    \textit{\# Graph attention block (Line 8-12)}
    \STATE $\mathbf{f}_\mathrm{node} = \mathrm{concat}_\mathrm{column-wise}(\mathbf{f}_\mathrm{atom}, \mathbf{f}_\mathrm{frame})$
    \STATE $\mathbf{f}_\mathrm{edge} = \mathrm{concat}_\mathrm{column-wise}(\mathrm{concat}_\mathrm{row-wise}(\mathbf{f}_\mathrm{aa}, \mathbf{f}_\mathrm{fa}^\top), \mathrm{concat}_\mathrm{row-wise}(\mathbf{f}_\mathrm{fa}, \mathbf{S}))$
    \STATE $\mathbf{X} = \mathrm{concat}_\mathrm{column-wise}(\mathrm{concat}_\mathrm{row-wise}(\mathbf{X}_\mathrm{a}, \mathbf{1}), \mathrm{concat}_\mathrm{row-wise}(\mathbf{1}, \mathbf{X}_\mathrm{s}))$
    \STATE $\mathbf{f}_\mathrm{out}, \_, \mathbf{z}_\mathrm{out} = \mathrm{MHAwithEdgeBias}(\mathbf{f}_\mathrm{node}, \mathbf{f}_\mathrm{node}, \mathbf{f}_\mathrm{edge}, \mathbf{X}, n_\mathrm{heads}=8, c_\mathrm{head}=8)$ \\
    \STATE $\mathbf{f}_\mathrm{edge} \gets \mathrm{MLP}(\mathrm{LinearNoBias}(\mathbf{z}_\mathrm{out}) + \mathbf{f}_\mathrm{edge}) + \mathbf{f}_\mathrm{edge}$ \\
    \textit{\# Note update block}
    \STATE $\mathbf{f}_\mathrm{node} \gets \mathrm{MLP}_{\mathrm{hidden\_dim}=2048}(\mathbf{f}_\mathrm{out} + \mathbf{f}_\mathrm{node}) + \mathbf{f}_\mathrm{node}$ 
    \STATE Update $\mathbf{f}_\mathrm{frame}$, $\mathbf{f}_\mathrm{fa}$ from merged embeddings $\mathbf{f}_\mathrm{node}$, $\mathbf{f}_\mathrm{edge}$
\ENDFOR
\IF{Subsample ligand pair representations}
    \STATE $\mathbf{F}_\mathrm{L} \gets \frac{1}{2}\big( \mathbf{H}_\mathrm{L} \cdot \mathbf{f}_\mathrm{fa} + (\mathbf{H}_\mathrm{L} \cdot \mathbf{f}_\mathrm{fa})^\top \big)$  \COMMENT{$\mathbf{F}_\mathrm{L} \in \mathbb{R}^{N_\mathrm{L} \times N_\mathrm{L} \times c_\mathrm{p}}$}
\ENDIF
\RETURN{} $\mathbf{f}_\mathrm{atom}, \mathbf{f}_\mathrm{frame}, \mathbf{f}_\mathrm{fa}, \mathbf{f}_\mathrm{aa}, \mathbf{F}_\mathrm{L}$
\end{algorithmic}
\end{algorithm}

$\mathrm{MHAwithEdgeBias}$ (Algorithm~\ref{alg:MHAwithEdgeBias}) denotes a multi-head cross-attention layer between source node embeddings and destination node embeddings, with edge embeddings entering attention computation as a relative positional encoding term. %
\begin{algorithm}[H]
\caption{Multi-head Graph Attention with Edge Bias.  $\mathbf{X}$ denotes the adjacency matrix of all edges on the graph.}
\label{alg:MHAwithEdgeBias}
\textbf{def} \text{MHAwithEdgeBias}($\mathbf{f}_\mathrm{src}, \mathbf{f}_\mathrm{dst}, \mathbf{f}_\mathrm{edge}, \mathbf{X}, c=512, c_\mathrm{p}=64, n_\mathrm{head}, c_\mathrm{head}$) \hfill {$\mathbf{f}_\mathrm{src}\in\mathbb{R}^{N_\mathrm{src} \times c}, \mathbf{f}_\mathrm{dst}\in\mathbb{R}^{N_\mathrm{dst} \times c}, \mathbf{f}_\mathrm{edge}\in\mathbb{R}^{N_\mathrm{edges} \times c_\mathrm{p}}$} \\
\begin{algorithmic}[1]
\STATE $\mathbf{f}_\mathrm{K}, \mathbf{f}_\mathrm{Vs} = \mathrm{LinearNoBias}(\mathbf{f}_\mathrm{src})$ \COMMENT{$\mathbf{f}_\mathrm{K}, \mathbf{f}_\mathrm{Vs} \in\mathbb{R}^{N_\mathrm{src} \times n_\mathrm{heads} \times c_\mathrm{heads}}$} 
\STATE $\mathbf{f}_\mathrm{Q}, \mathbf{f}_\mathrm{Vd} = \mathrm{LinearNoBias}(\mathbf{f}_\mathrm{dst}) $ \COMMENT{$\mathbf{f}_\mathrm{Q}, \mathbf{f}_\mathrm{Vd} \in\mathbb{R}^{N_\mathrm{dst} \times n_\mathrm{heads} \times c_\mathrm{heads}}$} 
\STATE $\mathbf{b} = \mathrm{LinearNoBias}(\mathbf{f}_\mathrm{edge}) $ \COMMENT{$\mathbf{b} \in\mathbb{R}^{N_\mathrm{edges} \times n_\mathrm{heads} \times 1}$} 
\STATE $\mathbf{z}_{ij} = \frac{1}{\sqrt{c_\mathrm{head}}}(\mathbf{f}_{\mathrm{K}, i} \cdot \mathbf{f}_{\mathrm{Q}, j}^\top) + \mathbf{b}_{ij} $ \COMMENT{$\mathbf{z} \in\mathbb{R}^{N_\mathrm{edges} \times n_\mathrm{heads} \times 1}$} 
\STATE $\mathbf{a}_{ij} = \mathrm{Softmax}_{j}^{X_{ij}=1} \, \mathbf{z}_{ij}$  \COMMENT{$\mathbf{a} \in\mathbb{R}^{N_\mathrm{edges} \times n_\mathrm{heads} \times 1}$} 
\STATE $\mathbf{f}_{\mathrm{os}, i} \gets \mathrm{LinearNoBias}(\sum_{j}^{X_{ij}=1}\mathbf{a}_{ij} \odot \mathbf{f}_{\mathrm{Vd}, j})$ \COMMENT{$\mathbf{f}_{\mathrm{os}} \in\mathbb{R}^{N_\mathrm{src} \times c}$} 
\STATE $\mathbf{f}_{\mathrm{od}, j} \gets \mathrm{LinearNoBias}(\sum_{j}^{X_{ij}=1}\mathbf{a}_{ij} \odot \mathbf{f}_{\mathrm{Vs}, i})$ \COMMENT{$\mathbf{f}_{\mathrm{od}} \in\mathbb{R}^{N_\mathrm{dst} \times c}$} 
\RETURN{} $\mathbf{f}_\mathrm{os}, \mathbf{f}_\mathrm{od}, \mathbf{z}$
\end{algorithmic}
\end{algorithm}

\begin{algorithm}
\caption{rigidFrom3Points (adapted from Ref.~\cite{jumper2021highly}, Alg. 21)}
\label{alg:rigidFrom3Points}
\textbf{def} \text{rigidFrom3Points}($\mathbf{x}_1\in \mathbb{R}^3, \mathbf{x}_2\in \mathbb{R}^3, \mathbf{x}_3\in \mathbb{R}^3, \epsilon=\SI{0.01}{\angstrom}$)
\begin{algorithmic}[1]
\STATE $\mathbf{v}_1 \gets \mathbf{x}_3 - \mathbf{x}_2$
\STATE $\mathbf{v}_2 \gets \mathbf{x}_1 - \mathbf{x}_2$
\STATE $\mathbf{e}_1 \gets \frac{\mathbf{v}_1}{\sqrt{\|\mathbf{v}_1\|_2^2 + \epsilon^2}}$
\STATE $\mathbf{u}_2 \gets \mathbf{v}_2 - (\mathbf{e}_1^\top \mathbf{v}_2) \mathbf{e}_1$
\STATE $\mathbf{e}_2 \gets \frac{\mathbf{u}_2}{\sqrt{\|\mathbf{u}_2\|_2^2 + \epsilon^2}}$
\STATE $\mathbf{e}_3 \gets \mathbf{e}_1 \times \mathbf{e}_2$
\STATE $\mathbf{R} \gets \text{vstack}(\mathbf{e}_1, \mathbf{e}_2, \mathbf{e}_3)$ \COMMENT{$\mathbf{R} \in \mathbb{R}^{3\times3}$}
\STATE $\mathbf{t} \gets \mathbf{x}_2$
\RETURN $\mathbf{t}, \mathbf{R}$
\end{algorithmic}
\end{algorithm}

\subsection{MHT model pretraining}

\label{sec:small_mol_pretraining}

\begin{table}[H]
\caption{Composition of the dataset used for pretraining the MHT chemical encoder.}
\label{table:small_mol_pretraining}
\centering
\begin{tabular}{@{}cccccc@{}}
\toprule
Data source & Num. samples collected & Sampling weight & $\mathcal{L}_\mathrm{3D}$ & $\mathcal{L}_\mathrm{REG}$ & $\mathcal{L}_\mathrm{MLM}$ \\ \midrule
 \makecell{BioLip~\cite{yang2012biolip} ligands \\ (deposited date\textless{}2019.1.1)} & 160k                   & 1.0             & +     & -     & +      \\
GEOM~\cite{axelrod2022geom} & 450k * 5               & 0.25             & +     & -     & +      \\
PEPCONF~\cite{prasad2019pepconf} & 3775                   & 5.0             & +     & -     & +      \\
PubChemQC~\cite{nakata2017pubchemqc,hu2021ogblsc} & 3.4M                   & 0.25             & +     & -     & +      \\
Chemical Checker~\cite{duran2020extending} & 800k                   & 1.0             & -     & +     & +      \\ \bottomrule
\end{tabular}
\end{table}

In Table~\ref{table:small_mol_pretraining} we summarize the small molecule datasets used for training the MHT encoder used in the reported NeuralPLexer model. The loss function used in MHT pretraining is the following:
\begin{equation}
    \mathcal{L}_\mathrm{lig-pretraining} = \mathcal{L}_\mathrm{3D-marginal} + \mathcal{L}_\mathrm{3D-DSM} + \mathcal{L}_\mathrm{CC-regression} + 0.01 \cdot \mathcal{L}_\mathrm{CC-ismask} + 0.1 \cdot \mathcal{L}_\mathrm{MLM}
\end{equation}

We use a mixture density network head to encourage alignment between the learned last-layer pair representations $\mathbf{G}$ and the intra-molecular 3D coordinate marginals. For a single training sample with 3D coordinate observation $\mathbf{y}$:
\begin{equation}
    \mathcal{L}_\mathrm{3D-marginal} = \sum_{u}^{N_\mathrm{frame}} \sum_{i}^{N_\mathrm{atom}} \log\big[ \sum_{l}^{N_\mathrm{modes}} \frac{ \exp(w_{iul}) \cdot q_\mathrm{3D}( T_u^{-1} \circ \mathbf{y}_i  |  \mathbf{m}_{iul})}{ \sum_{l}^{N_\mathrm{modes}} \exp(w_{iul})} \big]
\end{equation}
where $T_u\defeq (\mathbf{R}_u, \mathbf{t}_u)$, $T_u^{-1} \circ \mathbf{y}_i \defeq (\mathbf{y}_i - \mathbf{t}_u ) \cdot \mathbf{R}_u^\top$. 
$\mathbf{t}_u\in \mathbb{R}^3$ and $\mathbf{R}_u\in \mathrm{SO(3)}$ are given by 
\begin{equation}
    (\mathbf{R}_u, \mathbf{t}_u)= \mathrm{rigidFrom3Points}(\mathbf{y}_{i(u)}, \mathbf{y}_{j(u)}, \mathbf{y}_{k(u)})
\end{equation}
where $\mathrm{rigidFrom3Points}$ (Alg.~\ref{alg:rigidFrom3Points}) is the Gram–Schmidt-based frame construction operation originally described in Ref.~\cite{jumper2021highly}; we additionally add a numerical stability factor of $\SI{0.01}{\angstrom}$ to the vector-norm calculations to handle edge cases when computing the rotation matrices from perturbed coordinates.
Each component the 3D distance-angle distribution $q^\mathrm{3D}$ is parameterized by
\begin{equation}
    q_\mathrm{3D}(\mathbf{t} | \mu, \sigma, \mathbf{v}) = \mathrm{Gaussian}(\lVert \mathbf{t} \rVert_2 | \mu, \sigma) \times \mathrm{PowerSpherical}(\frac{\mathbf{t}}{\lVert \mathbf{t} \rVert_2}| \mathbf{v}, d=3) 
\end{equation}
where $\mathrm{PowerSpherical}$ is a power spherical distribution introduced in~\cite{de2020power}; $\mathbf{m}_{iul} \defeq (\mu, \sigma, \mathbf{v})_{iul}$, and
\begin{equation}
    [\mathbf{w}_{iu}, \mathbf{m}_{iu}] = \mathrm{3DMixtureDensityHead}\big(\mathbf{G}_{l_\mathrm{max}} \big)_{iu} .
\end{equation}
where $\mathrm{3DMixtureDensityHead}$ is a 3-layer MLP.

Using an equivariant graph transformer similar to ESDM (see Sec.~\ref{sec:ESDM}) but with all receptor nodes dropped, we construct a geometry prediction head to perform global molecular 3D structure denoising. We sample noised coordinates $\mathbf{y}(t)$ from a VPSDE~\cite{song2020score} and introduce a $\mathrm{SE(3)}$-invariant denoising score matching loss based on the Frame Aligned Point Error (FAPE)~\cite{jumper2021highly}:
\begin{equation}
    \mathcal{L}_\mathrm{3D-DSM} = \mathbb{E}_{t\sim (0, 1], \mathbf{y}_t \sim q_{0:t}(\cdot | \mathbf{y})} \big[ \mathrm{mean}_{u,i} \min( \lVert T_u^{-1} \circ \mathbf{y}_i - \hat{T}_u^{-1} \circ \hat{\mathbf{y}}_i \rVert_2, \SI{10}{\angstrom}) \cdot \sqrt{\alpha_t} \big]
\end{equation}
where
\begin{equation}
    \hat{\mathbf{y}} = \mathrm{GeometryPredictionHead}(\mathbf{y}_t; \mathbf{H}_{l_\mathrm{max}}, \mathbf{F}_{l_\mathrm{max}}, \mathbf{G}_{l_\mathrm{max}})
\end{equation}

$\mathcal{L}_\mathrm{CC-regression}$ is a SmoothL1 loss for fitting the "level 1" chemical checker (CC)~\cite{duran2020extending} embeddings which represents harmonized and integrated bioactivity data, and $\mathcal{L}_\mathrm{CC-ismask}$ is an auxiliary binary cross entropy loss for classifying whether a specific CC entry is available for any molecule in the chemical checker dataset. $\mathcal{L}_\mathrm{MLM}$ is a standard cross-entropy loss for predicting the masked tokens. %
The MHT model is trained with a $50\%$ masking ratio for all the atom, bond, edge, and stereochemistry encodings, and with dropout=0.1; we trained the model with batch size = 32 for $1.5 \times 10^6$ iterations using a cosine annealing schedule, taking 184 hours on a single NVIDIA-Tesla-V100-SXM2-32GB GPU.

\clearpage

\section{Protein-ligand graph featurization}

\label{sec:graph_construction}

\begin{table}[h]
\centering
\caption{Notations for all node types in the protein-ligand graph (also see Figure~\ref{fig:fig2}e).}
\label{table:all_nodes}
\begin{tabular}{llllll}
\toprule
Abbr. & Count & Meaning & Initial features \\
\midrule
B & $N_\mathrm{res}$ & Residue-wise backbone frames & PLM + template node features (see Sec.~\ref{sec:protfeats}) \\
S & $N_\mathrm{P}$ & Patch-wise anchor frames & \textit{subset of above (B)} \\
F & $N_\mathrm{L}$ & Ligand anchor frames & MHT frame node embeddings $\mathbf{f}_\mathrm{frame}$ \\
P & $N_\mathrm{protatm}$ & All protein atoms & MHT atomic node embeddings $\mathbf{f}_\mathrm{atom}$ \\
L & $N_\mathrm{ligatm}$ & All ligand atoms & MHT atomic node embeddings $\mathbf{f}_\mathrm{atom}$ \\
\bottomrule
\end{tabular}
\end{table}

\begin{table}[h]
\centering
\caption{All edge types in the protein-ligand graph (also see Figure~\ref{fig:fig2}e). All edges comprised of two distinct node types are bidirectional; for conciseness, only one of the two directions is explicitly shown below. $\mathrm{RBF}(\cdot)$ denotes a damped random Fourier basis function layer as defined in Eq.~\eqref{eq:RBF}. For the exponential-kNN scheme, see Algorithm~\ref{alg:exponentialknn}.}
\label{table:all_edges}
\begin{tabular}{llllll}
\toprule
Abbr. & Edge type & $src$ type & $dst$ type & Connectivity & Initial features \\
\midrule
\multicolumn{5}{l}{\# \textit{Residue-scale subgraph}} \\ 
BB & local & B & B & Exponential-kNN($\mathbf{x}_\mathrm{C\alpha}$), $k_\mathrm{NN}=32$ & See Section~\ref{sec:protfeats} \\
BS & local & B & S & Intra-patch, dense & See Section~\ref{sec:protfeats} \\
BF & long-range & B & F & Dense & Node embedding outer sum \\
SS & long-range & S & S & Dense  & See Section~\ref{sec:protfeats} \\
SF & long-range & S & F & Dense  & Zero tensors \\
FF & long-range & F & F & Dense  & Symmetrized MHT embeddings $\mathbf{F}_\mathrm{L}$  \\
\multicolumn{5}{l}{\# \textit{Atomic-scale subgraph}} \\ 
As & local & P+L & P+L & Exponential-kNN($\mathbf{Z}$), $k_\mathrm{NN}=8$ & $\mathrm{RBF}_{32}(\lVert \mathbf{Z}_{src} - \mathbf{Z}_{dst} \rVert / \SI{10}{\angstrom})$ \\
Ac & local & P+L & P+L & Molecular graph n-hop, n=3 & Gathered MHT embeddings $\mathbf{f}_\mathrm{aa}$ \\
Ab & local & P+L & P+L & Inter-molecular covalent bonds & Trainable random embedding \\
\multicolumn{5}{l}{\# \textit{Cross-scale edges connecting atoms and residues}} \\ 
BP & local & B & P & Intra-residue, dense & Gathered MHT embeddings $\mathbf{f}_\mathrm{fa}$ \\
FL & local & F & L & Intra-ligand, dense & Gathered MHT embeddings $\mathbf{f}_\mathrm{fa}$ \\
\bottomrule
\end{tabular}
\end{table}

Given the primary model inputs and a noisy geometry, the schemes for constructing the residue-scale and atomic-scale graph representations are summarized in Table~\ref{table:all_nodes}-\ref{table:all_edges}. 
The protein anchor frame nodes (Table~\ref{table:all_nodes} S) are selected by first sequentially segmenting the input protein sequence into $N_\mathrm{P}=96$ patches of the same sequence length (with the last patch potentially truncated), and then sampling one unique backbone frame index for each protein patch. The intra-patch edges (Table~\ref{table:all_edges} BS) then connect each protein residue node to the anchor node within the same patch.

The geometry-dependent local edges (Table~\ref{table:all_edges} BB, As) are generated using a randomized k-nearest-neighbor (kNN) scheme with an exponentially-decaying attachment probability $p(\mathrm{add\_edge}(i, j)) \propto \exp(- \lVert \mathbf{Z}_i - \mathbf{Z}_j \rVert / \SI{5.0}{\angstrom})$ with respect to the distance matrix among nodes, implemented via a Gumbel-Topk trick (note that a similar scheme is adopted in Ref.~\cite{Ingraham2022.12.01.518682}):
\begin{algorithm}[H]
\caption{Exponential-kNN scheme for generating local edges}
\label{alg:exponentialknn}
\begin{algorithmic}[1]
\REQUIRE All node Euclidean coordinates $\mathbf{Z}$, node degree $k_\mathrm{NN}$, decay scale $D=\SI{5.0}{\angstrom}$
\FOR{$i \in \{1, \cdots, N_\mathrm{nodes}\}$}
    \FOR{$j \in \{1, \cdots, N_\mathrm{nodes}\}$}
        \STATE $s_{ij} \gets - \lVert \mathbf{Z}_i - \mathbf{Z}_j \rVert / D$
        \STATE $u_{ij} \sim \mathrm{Uniform}(0, 1)$
        \STATE $\tilde{s}_{ij} \gets s_{ij} - \log( -\log( u_{ij} ) ) $ \COMMENT{Gumbel distribution over logits $s_{ij}$}
    \ENDFOR
    \STATE $\{src\_idx\}_i, \{dst\_idx\}_i \gets \mathrm{ArgTopK}_{j}(\{\tilde{s}_{ij}\}_i), i$ \COMMENT{Row-wise top-k node indices, k=$k_\mathrm{NN}$}
\ENDFOR
\RETURN $\{src\_idx\}, \{dst\_idx\}$
\end{algorithmic}
\end{algorithm}

$\mathrm{RBF}(\cdot): \mathbb{R} \rightarrow \mathbb{R}^{2 * N_\mathrm{basis}}$ denotes a damped random Fourier Feature encoding layer with sine and cosine frequencies $\bm{\kappa} \in \mathbb{R}^{N_\mathrm{basis}}$ initialized from a univariate Gaussian distribution:
\begin{equation}
    \label{eq:RBF}
    \mathrm{RBF}_{N_\mathrm{basis}}(r) \defeq [\sin(\bm{\kappa} \cdot r), \cos(\bm{\kappa} \cdot r)]^\top / (1 + r)
\end{equation}

The inter-molecular covalent edges (Table~\ref{table:all_edges} Ab, such as those in post-translational modifications and polysaccharides for which monomers are deposited as individual ligands) are determined based on the reference complex structure; an atom pair $(i, j)$ is considered a covalent bond iff $d_{ij} < 1.2 \sigma_{ij}$ where $\sigma_{ij} = \frac{1}{2}(\sigma_i + \sigma_j)$ is the average Van der Waals (VdW) radius.

\subsection{Protein residue sub-graph featurization}

\label{sec:protfeats}

The initial node features of all protein residue nodes are a concatenation of (a) the one-hot amino-acid types (20 standard residues + 1 "unknown" token) %
and (b) the concatenation of ESM-2-650M embeddings~\cite{lin2023evolutionary} computed for all input protein sequences, and, when available, (c) internal coordinates of all atoms in the template protein structure $\mathbf{x}_\mathrm{template}$ in the corresponding backbone coordinate frame, padded into the fixed-length atom37 format of PDB amino acid atom types. The initial node features are then embedded by a standard 3-layer MLP: $\mathbb{R}^{N_\mathrm{res} \times (1280+21+37*3)} \rightarrow \mathbb{R}^{N_\mathrm{res} \times c}$. 

As detailed in Algorithm~\ref{alg:protedgefeat}, the initial edge features of all protein residue-residue edges (Table~\ref{table:all_edges} BB, BS, SS) are a combination of (a) an outer-sum of source and destination node features, (b) relative positional encodings of residue indices in the protein sequences, (c) relative geometrical encodings of residue backbones in the input noisy protein structure $\mathbf{x}_t$, and, when available, (d) relative geometrical encodings of residue backbones in the template protein structure $\mathbf{x}_\mathrm{template}$.

\begin{algorithm}
\caption{Computing the initial feature for a single residue-residue edge}
\label{alg:protedgefeat}
\begin{algorithmic}[1]
\REQUIRE $src\_idx$, $dst\_idx$, residue node features $\mathbf{f}_\mathrm{B}$, input protein coordinates $\mathbf{x}_t$, input template coordinates $\mathbf{x}_\mathrm{template}$ (optional)
\STATE $\mathbf{f}_\mathrm{osum} = (\mathbf{f}_\mathrm{B})_{src\_idx} + (\mathbf{f}_\mathrm{B})_{dst\_idx} $ \COMMENT{Eq.~\ref{eq:RBF}}
\STATE $\mathbf{f}_\mathrm{seq} = \mathrm{RBF}_{16}(residue\_idx(src\_idx) - residue\_idx(dst\_idx)) \cdot \mathrm{IsSameChain}(src\_idx, dst\_idx)$
\STATE $\mathbf{f}_\mathrm{geom} = \mathrm{RelGeomEnc}(src\_idx, dst\_idx, \mathbf{x}_t, N_\mathrm{basis}=15)$ \COMMENT{Alg.~\ref{alg:relgeoenc}}
\STATE $\mathbf{f}_\mathrm{edge} \gets \mathrm{MLP}(\mathrm{concat}(\mathbf{f}_\mathrm{osum}, \mathbf{f}_\mathrm{seq}, \mathbf{f}_\mathrm{geom}))$
\IF{$\mathbf{x}_\mathrm{template}$ is not None}
    \STATE $\mathbf{f}_\mathrm{edge} \gets \mathbf{f}_\mathrm{edge} + \mathrm{LinearNoBias}(\mathrm{RelGeomEnc}(src\_idx, dst\_idx, \mathbf{x}_\mathrm{template}, N_\mathrm{basis}=15) )$ \COMMENT{Alg.~\ref{alg:relgeoenc}}
\ENDIF
\RETURN $\mathbf{f}_\mathrm{edge}$
\end{algorithmic}
\end{algorithm}

\begin{algorithm}
\caption{Computing the relative geometrical encodings for a single residue-residue edge}
\label{alg:relgeoenc}
\textbf{def} {RelGeomEnc}($src\_idx, dst\_idx, \mathbf{x}, N_\mathrm{basis}, D_0=\SI{10.0}{\angstrom}$)
\begin{algorithmic}[1]
\STATE $\{\mathbf{x}_\mathrm{N}, \mathbf{x}_\mathrm{C\alpha}, \mathbf{x}_\mathrm{C}\} \gets$ Get protein backbone (N, C$\mathrm{\alpha}$, C) coordinates from $\mathbf{x}$
\STATE $\mathbf{t}_\mathrm{src}, \mathbf{R}_\mathrm{src} = \mathrm{rigidFrom3Points}((\mathbf{x}_\mathrm{N})_{src\_idx}, (\mathbf{x}_\mathrm{C\alpha})_{src\_idx}, (\mathbf{x}_\mathrm{C})_{src\_idx})$ \COMMENT{Alg.~\ref{alg:rigidFrom3Points}}
\STATE $\mathbf{t}_\mathrm{dst}, \mathbf{R}_\mathrm{dst} = \mathrm{rigidFrom3Points}((\mathbf{x}_\mathrm{N})_{dst\_idx}, (\mathbf{x}_\mathrm{C\alpha})_{dst\_idx}, (\mathbf{x}_\mathrm{C})_{dst\_idx})$ \COMMENT{Alg.~\ref{alg:rigidFrom3Points}}
\STATE $d = \lVert \mathbf{t}_\mathrm{src} - \mathbf{t}_\mathrm{dst} \rVert $
\STATE $\mathbf{f}_\mathrm{dist} = \mathrm{RBF}_{N_\mathrm{basis}}(d / D_0)$
\STATE $\mathbf{f}_\mathrm{dirs} = \mathbf{R}_\mathrm{src}^\top \cdot (\mathbf{t}_\mathrm{dst} - \mathbf{t}_\mathrm{src}) / (d + \SI{1.0}{\angstrom})$
\STATE $\mathbf{f}_\mathrm{dird} = \mathbf{R}_\mathrm{dst}^\top \cdot (\mathbf{t}_\mathrm{src} - \mathbf{t}_\mathrm{dst}) / (d + \SI{1.0}{\angstrom})$
\STATE $\mathbf{f}_\mathrm{ori} = \mathrm{flatten}(\mathbf{R}_\mathrm{src}^\top \cdot \mathbf{R}_\mathrm{dst})$
\STATE $\mathbf{f}_\mathrm{geom} = \mathrm{concat}(\mathbf{f}_\mathrm{dist}, \mathbf{f}_\mathrm{dirs}, \mathbf{f}_\mathrm{dird}, \mathbf{f}_\mathrm{ori})$ \COMMENT{$\mathbf{f}_\mathrm{geom} \in \mathbb{R}^{N_\mathrm{basis}+15}$}
\RETURN $\mathbf{f}_\mathrm{geom}$
\end{algorithmic}
\end{algorithm}

\section{The CPM network architecture}
\label{sec:CPM}

\begin{sidewaysfigure}
    \centering
    \includegraphics[width=\linewidth]{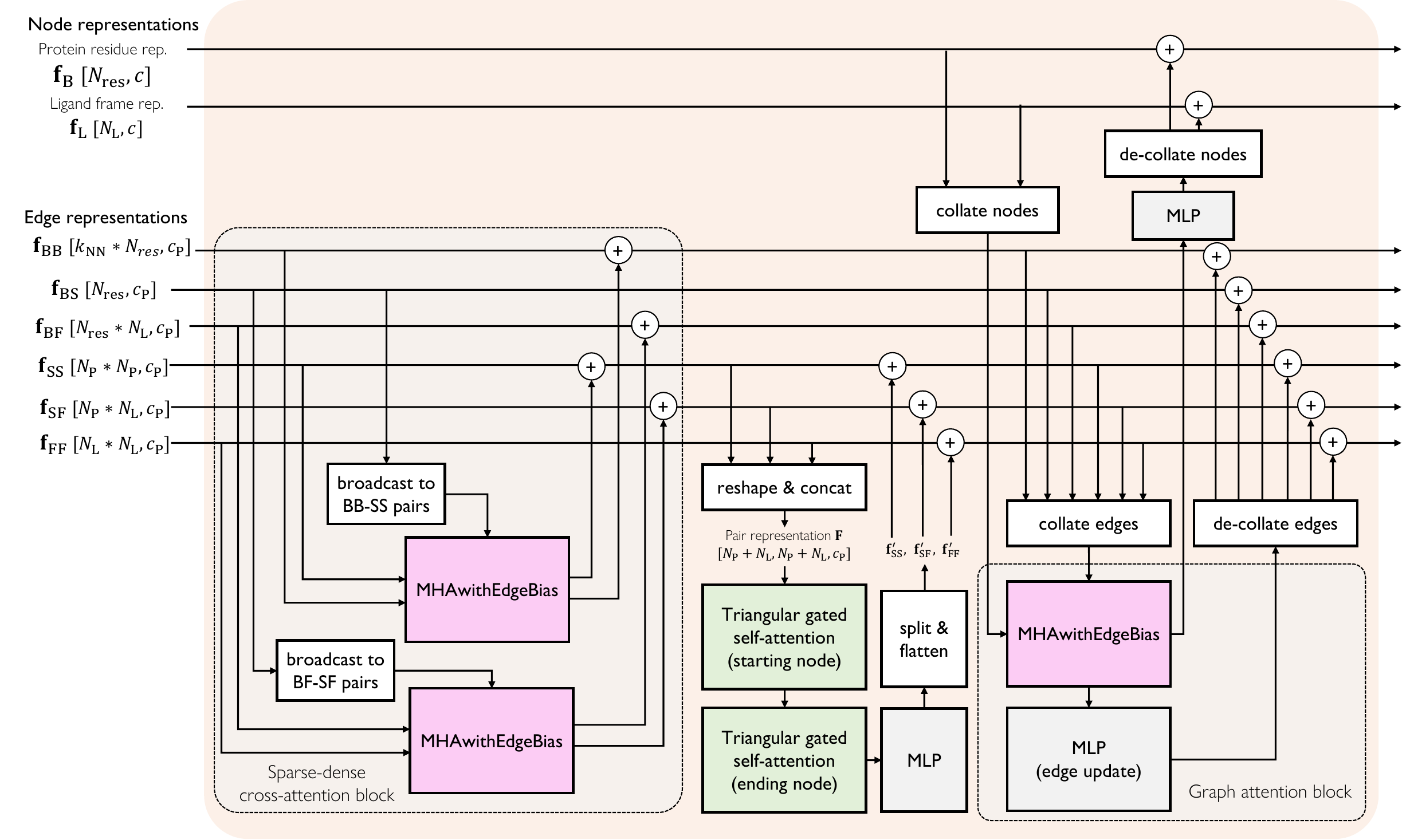}
    \caption{Network architecture of a single contact prediction module (CPM) block. Arrows indicate information flow directions, and "+" indicates an element-wise tensor summation. The Triangular gated self-attention (around starting/ending node) blocks refer to Algorithm 13-14 of Ref.\cite{jumper2021highly}. $k_\mathrm{NN}$ denotes the neighbor size of local residue-residue edges (see Section~\ref{sec:graph_construction}), and we use $k_\mathrm{NN}=32$ for all models reported in this work. Note that $\mathbf{f}_\mathrm{SS} \equiv \mathbf{F}_\mathrm{P}$ and $\mathbf{f}_\mathrm{FF} \equiv \mathbf{F}_\mathrm{L}$ up to a tensor reshaping operation.}
    \label{fig:cpm}
\end{sidewaysfigure}

The neural network architecture of a single contact prediction module (CPM) block is summarized in Figure~\ref{fig:cpm}; for all models reported in this work, we use a stack of $N_\mathrm{CPM}=6$ blocks to construct the network (i.e., CPMForward in Algorithm~\ref{alg:main}). 

Before the CPM forward pass, all frame node representations are concatenated with a 64-bit random Fourier encoding of the diffusion time $\mathrm{RBF}(\tau)$ which is embedded by a standard $\mathrm{MLP}$. The last step sampled block-adjacency matrix $\tilde{\mathbf{l}}$ is encoded by a $\mathrm{LinearNoBias}$ layer; the block-adjacency encodings are then added to the edge representations between patch-wise protein nodes and selected ligand nodes $\mathbf{f}_\mathrm{SF} \in \mathbb{R}^{N_\mathrm{P} * N_\mathrm{L} \times c_\mathrm{P}}$.

The first 2 CPM blocks are executed on the protein sub-graph only (i.e., BB, BS, and SS edges as defined in Table~\ref{table:all_edges}), and the remaining 4 CPM blocks are executed on the entire residue-scale graph (i.e., BB, BS, BF, SS, SF, and FF edges as defined in Table~\ref{table:all_edges}).

The asymptotic computational complexity of $\mathrm{CPMForward}$ is $\mathcal{O}\big(N_\mathrm{CPM} \cdot ( N_\mathrm{res} \cdot (k_\mathrm{NN} + N_\mathrm{L}) + (N_\mathrm{P} + N_\mathrm{L})^3 ) \big)$.

\section{The ESDM network architecture}
\label{sec:ESDM}

The Equivariant Structure Denoising Module (ESDM) of the NeuralPLexer network predicts denoised three-dimensional structures $\hat{\mathbf{Z}}_0$ using the noise input coordinates $\mathbf{Z}_t$ and graph representations of the binding complex.
The neural network architecture of a single ESDM block is summarized in Figure~\ref{fig:esdm}; for all models reported in this work, we use a stack of $N_\mathrm{ESDM}=4$ blocks to construct the network (i.e., ESDMForward in Algorithm~\ref{alg:main}). 

\begin{figure}
    \centering
    \includegraphics[width=\textwidth]{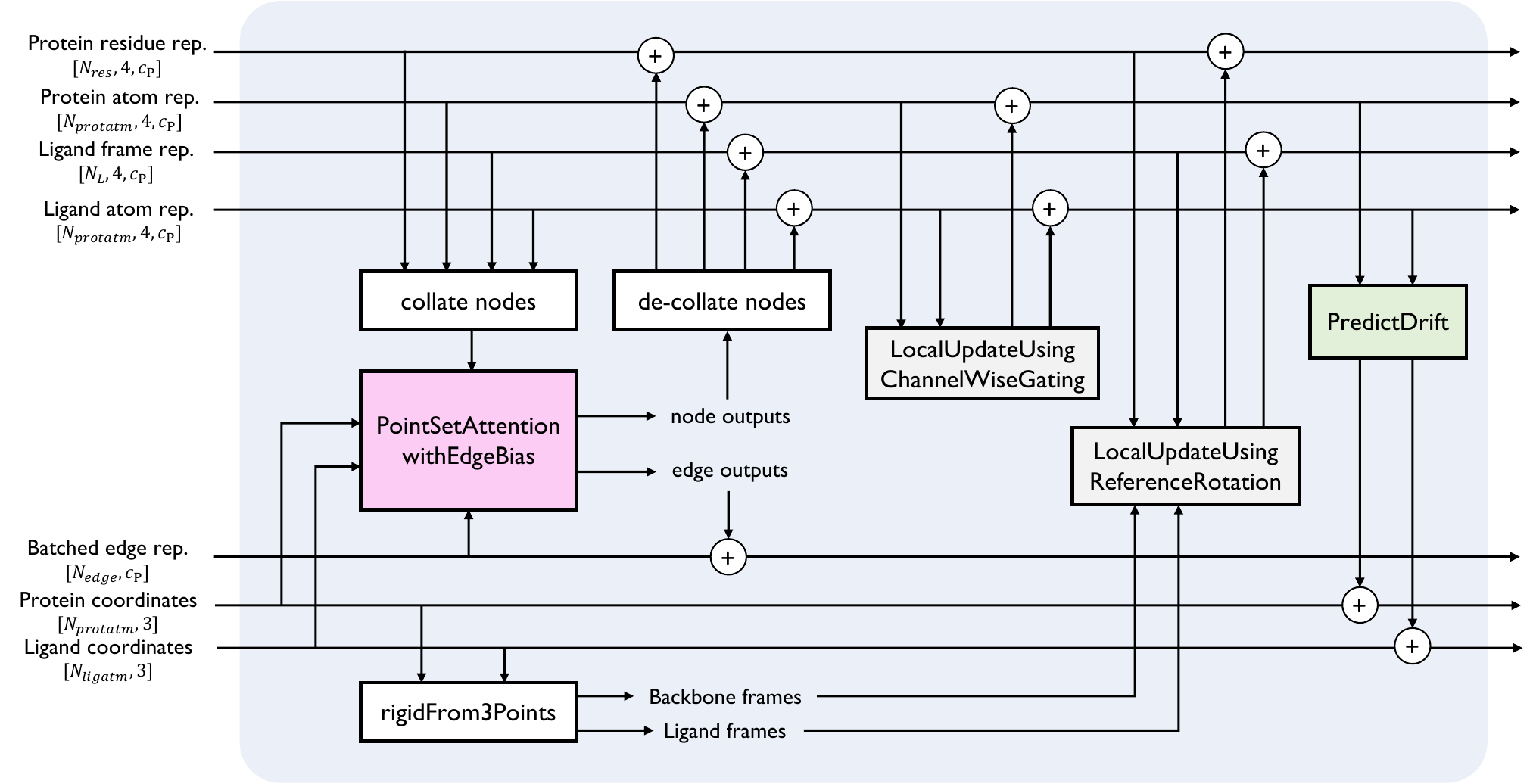}
    \caption{Network architecture of a single block in the equivariant structure diffusion module (ESDM). Arrows indicate information flow directions, and "+" indicates an element-wise tensor summation.}
    \label{fig:esdm}
\end{figure}

Before the ESDM forward pass, all atomic node representations are concatenated with a 64-bit random Fourier encoding of the diffusion time $\mathrm{RBF}(\tau)$ which is embedded by a standard $\mathrm{MLP}$. The forward pass expressions of trainable modules $\mathrm{PointSetAttentionwithEdgeBias}$, $\mathrm{LocalUpdateUsingChannelWiseGating}$, $\mathrm{LocalUpdateUsingReferenceRotation}$, $\mathrm{PredictDrift}$ are defined as:
\begin{algorithm}[H]
\caption{PointSetAttentionwithEdgeBias}
\label{alg:PointSetAttentionwithEdgeBias}
\textbf{def} {PointSetAttentionwithEdgeBias}($\mathbf{f}_\mathrm{s}, \mathbf{f}_\mathrm{v}, \mathbf{f}_\mathrm{e}, \mathbf{t}$, $c=64$) \hfill $\mathbf{f}_\mathrm{s}\in\mathbb{R}^{N_\mathrm{nodes} \times c},  \mathbf{f}_\mathrm{v}\in\mathbb{R}^{N_\mathrm{nodes} \times 3 \times c}, \mathbf{f}_\mathrm{e}\in\mathbb{R}^{N_\mathrm{edges} \times c}, \mathbf{t}\in\mathbb{R}^{N_\mathrm{nodes} \times 3}$ \\
\begin{algorithmic}[1]
\STATE $\mathbf{f}_\mathrm{Q}, \mathbf{f}_\mathrm{K}, \mathbf{f}_\mathrm{V} = \mathrm{LinearNoBias}(\mathbf{f}_\mathrm{s})$ \COMMENT{$\mathbf{f}_\mathrm{Q}, \mathbf{f}_\mathrm{K}, \mathbf{f}_\mathrm{V} \in \mathbb{R}^{N_\mathrm{nodes} \times n_\mathrm{heads} \times c_\mathrm{head}}$}
\STATE $\mathbf{t}_\mathrm{Q}, \mathbf{t}_\mathrm{K}, \mathbf{t}_\mathrm{V} = \mathbf{t} / D_\mathrm{points} + \mathrm{LinearNoBias}(\mathbf{f}_\mathrm{v})$ \COMMENT{$\mathbf{t}_\mathrm{Q}, \mathbf{t}_\mathrm{K}, \mathbf{t}_\mathrm{V} \in \mathbb{R}^{N_\mathrm{nodes} \times n_\mathrm{heads} \times c_\mathrm{point} \times 3}, D_\mathrm{points}=\SI{10}{\angstrom}$}
\STATE $\mathbf{b} = \mathrm{LinearNoBias}(\mathbf{f}_\mathrm{e})$ \COMMENT{$\mathbf{b}\in\mathbb{R}^{N_\mathrm{edges} \times c_\mathrm{heads}}$} \\
\textit{\# Computing attention weights on all edges of the graph}
\STATE $\mathbf{z}_{ij} = \frac{1}{\sqrt{c_\mathrm{head}}}(\mathbf{f}_\mathrm{Q,i}^\top \cdot  \mathbf{f}_\mathrm{K,j}) + \mathbf{b}_{ij} - \frac{\mathbf{w}_{ij}}{\sqrt{18 c_\mathrm{head}}} \lVert \mathbf{t}_{\mathrm{Q},i} - \mathbf{t}_{\mathrm{K},j} \rVert_2^2$ \COMMENT{$\mathbf{z}\in \mathbb{R}^{N_\mathrm{edges} \times n_\mathrm{heads}}$}
\STATE $\bm{\alpha}_{ij} = \mathrm{Softmax}_{j\in \{i\}}(\mathbf{z}_{ij})$
\STATE $\mathbf{f}_\mathrm{s}' \gets \sum_{j\in \{i\}} \bm{\alpha}_{ij} \odot \mathbf{f}_\mathrm{V}$ 
\STATE $\mathbf{f}_\mathrm{v}' \gets (\sum_{j\in \{i\}} \bm{\alpha}_{ij} \odot \mathbf{t}_\mathrm{V}) - \mathbf{t} / D_\mathrm{points}$
\STATE $\mathbf{f}_\mathrm{e}' \gets \mathrm{MLP}(\mathbf{z}_{ij}) + \mathbf{f}_\mathrm{e}$
\RETURN $\mathbf{f}_\mathrm{s}', \mathbf{f}_\mathrm{v}', \mathbf{f}_\mathrm{e}'$
\end{algorithmic}
\end{algorithm}

The expression for computing attention weights $\mathbf{z}$ is adapted from Invariant Point Attention (IPA)~\cite{jumper2021highly}.

\begin{algorithm}[H]
\caption{LocalUpdateUsingChannelWiseGating}
\label{alg:LocalUpdateUsingChannelWiseGating}
\textbf{def} {LocalUpdateUsingChannelWiseGating}($\mathbf{f}_\mathrm{s}, \mathbf{f}_\mathrm{v}$) \hfill $\mathbf{f}_\mathrm{s}\in\mathbb{R}^{N_\mathrm{nodes} \times c},  \mathbf{f}_\mathrm{v}\in\mathbb{R}^{N_\mathrm{nodes} \times 3 \times c}$ \\
\begin{algorithmic}[1]
\STATE $\mathbf{f}_\mathrm{loc} = \mathrm{concat}(\mathbf{f}_\mathrm{s}, \lVert \mathbf{f}_\mathrm{v} \rVert_2)$ \COMMENT{$\mathbf{f}_\mathrm{loc}\in\mathbb{R}^{N_\mathrm{nodes} \times 2c}$}
\STATE $\mathbf{f}_\mathrm{s}', \mathbf{f}_\mathrm{gate} \gets \mathrm{MLP}(\mathbf{f}_\mathrm{loc})$ \COMMENT{$\mathbf{f}_\mathrm{gate}\in\mathbb{R}^{N_\mathrm{nodes} \times 1 \times c}$}
\STATE $\mathbf{f}_\mathrm{v}' \gets \mathrm{LinearNoBias}(\mathbf{f}_\mathrm{v}) \odot \mathbf{f}_\mathrm{gate}$ \COMMENT{Channel-wise product (broadcasting along \textit{xyz})}
\RETURN $\mathbf{f}_\mathrm{s}', \mathbf{f}_\mathrm{v}'$
\end{algorithmic}
\end{algorithm}
As only linear layers and vector scaling operations are used to update the vector representations $\mathbf{f}_\mathrm{v}$, \eqref{alg:LocalUpdateUsingChannelWiseGating} is $\mathrm{E(3)}$-equivariant.

\begin{algorithm}[H]
\caption{LocalUpdateUsingReferenceRotation}
\label{alg:LocalUpdateUsingReferenceRotation}
\textbf{def} {LocalUpdateUsingReferenceRotation}($\mathbf{f}_\mathrm{s}, \mathbf{f}_\mathrm{v}, \mathbf{R}$) \hfill $\mathbf{f}_\mathrm{s}\in\mathbb{R}^{N_\mathrm{nodes} \times c},  \mathbf{f}_\mathrm{v}\in\mathbb{R}^{N_\mathrm{nodes} \times 3 \times c}, \mathbf{R}\in \mathrm{SO(3)}^{N_\mathrm{nodes}}$ \\
\begin{algorithmic}[1]
\STATE $(\mathbf{f}_\mathrm{vloc})_i = (\mathbf{R}_i)^{\mathrm{T}} \cdot (\mathbf{f}_\mathrm{v})_i$ \COMMENT{$i\in\{1, \cdots, N_\mathrm{nodes}\}$} 
\STATE $\mathbf{f}_\mathrm{loc} = \mathrm{concat}(\mathbf{f}_\mathrm{s}, \mathbf{f}_\mathrm{vloc}, \lVert \mathbf{f}_\mathrm{v} \rVert_2)$ \COMMENT{$\mathbf{f}_\mathrm{loc}\in\mathbb{R}^{N_\mathrm{nodes} \times 5c}$}
\STATE $\mathbf{f}_\mathrm{s}', \mathbf{f}_\mathrm{vloc} \gets \mathrm{MLP}(\mathbf{f}_\mathrm{loc})$
\STATE $(\mathbf{f}_\mathrm{v}')_i \gets \mathbf{R}_i \cdot (\mathbf{f}_\mathrm{vloc})_i$ \COMMENT{$i\in\{1, \cdots, N_\mathrm{nodes}\}$} 
\RETURN $\mathbf{f}_\mathrm{s}', \mathbf{f}_\mathrm{v}'$
\end{algorithmic}
\end{algorithm}
Since the third row of $\mathbf{R}$ is a pseudovector as described in $\mathrm{rigidFrom3Points}$, the determinant of the rotation matrix $\mathbf{R}$ is unchanged under parity inversion transformations $i: \mathbf{x}\mapsto-\mathbf{x}$ and thus the intermediate quantity $\mathbf{f}_\mathrm{vloc}$ is  $\mathrm{SE(3)}$-invariant but in general \textbf{not invariant} under parity inversion $i$. This property ensures that ESDM-predicted coordinates can capture the correct chiral symmetry-breaking behaviors in molecular 3D conformation distributions.

\begin{algorithm}[H]
\caption{PredictDrift}
\label{alg:PredictDrift}
\textbf{def} {PredictDrift}($\mathbf{f}_\mathrm{s}, \mathbf{f}_\mathrm{v}$) \hfill $\mathbf{f}_\mathrm{s}\in\mathbb{R}^{N_\mathrm{nodes} \times c},  \mathbf{f}_\mathrm{v}\in\mathbb{R}^{N_\mathrm{nodes} \times 3 \times c}$ \\
\begin{algorithmic}[1]
\STATE $\mathbf{o}_\mathrm{scale} \gets \mathrm{Softplus}(\mathrm{MLP}(\mathbf{f}_\mathrm{s}))$ \COMMENT{$\mathbf{o}_\mathrm{scale}\in\mathbb{R}^{N_\mathrm{nodes} \times 1}$}
\STATE $\Delta \mathbf{t} \gets \mathrm{LinearNoBias}(\mathbf{f}_\mathrm{v}) \odot \mathbf{o}_\mathrm{scale}$ \COMMENT{$\Delta \mathbf{t} \in\mathbb{R}^{N_\mathrm{nodes} \times 3}$}
\RETURN $\Delta \mathbf{t}$
\end{algorithmic}
\end{algorithm}
The predicted drift vectors $\Delta \mathbf{t}$ are added to the atomic coordinates from the last ESDM block; after $N_\mathrm{ESDM}$ blocks, the final coordinate outputs are taken as the predicted denoised structure $\hat{\mathbf{Z}}_0$ to infer the score function.

\section{Confidence estimation heads}
\label{sec:si_plddt}

The predicted lDDT (pLDDT) head of NeuralPLexer uses the same architecture as ESDM (Sec.~\ref{sec:ESDM}) with an independent set of parameters. Once a structure $\mathbf{Z}_0$ is generated from the main network, $\mathbf{Z}_0$ and an empty block contact map is passed to the pre-trained CPM network to generate the residue-scale graph embeddings to the pLDDT head (Algorithm~\ref{alg:main}, lines 41-44). 

The output protein residue representations and ligand atom representations from the plDDT head are passed to a 6-layer $\mathrm{MLP}$ to predict the histograms of distance error distributions between generated and reference structures, that is, for each query protein residue centroid or ligand atom $i$, we fit the histogram of distance deviations $| \lVert (\mathbf{Z}_0)_i - (\mathbf{Z}_0)_j \rVert - \lVert (\mathbf{Z}_\mathrm{ref})_i - (\mathbf{Z}_\mathrm{ref})_j \rVert |$ for all target point  $j$ within \SI{15.0}{\angstrom} of the query point in the reference structure $\mathbf{Z}_\mathrm{ref}$. For such histograms, we use distance deviation bins with left boundaries of $[0.0, 0.5, 1.0, 2.0, 4.0, 8.0, 16.0]\SI{}{\angstrom}$. 
The pLDDT score of each protein residue or ligand atom is then computed based on the respective predicted histogram densities: 
\begin{equation}
    \label{eq:plddt}
    {pLDDT}[i] \defeq 1.0 * p_\mathrm{0-0.5}[i] + 0.75 * p_\mathrm{0.5-1.0}[i] + 0.5 * p_\mathrm{1.0-2.0}[i] + 0.25 * p_\mathrm{2.0-4.0}[i] .
\end{equation}

\clearpage

\section{Model training details}

\label{sec:si_training}

In Algorithm~\ref{alg:training} we summarize the general procedure for training the NeuralPLexer main network and confidence estimation heads.
$\mathrm{Bern}(p)$ denotes a Bernoulli distribution of a 0-1 boolean variable $X$ with $Pr(X=1)=p$. $\mathrm{bucketize\_onehot}(\cdot, \mathbf{v}_\mathrm{bins}))$ denotes a one-hot encoding operation using left bin boundaries defined in a vector $\mathbf{v}_\mathrm{bins}$. $\mathcal{L}_\mathrm{CE}(\cdot, \cdot)$ denotes a standard cross entropy loss between two discrete distributions: $\mathcal{L}_\mathrm{CE}(\mathbf{p}, \mathbf{q})\defeq - \sum_{i=1}^{N_\mathrm{bins}} \log{p_i} \cdot q_i$.

\begin{algorithm}
\caption{NeuralPLexer Training}
\label{alg:training}
\begin{algorithmic}[1]
\REQUIRE $\{\mathbf{s}\}$, $\{{G}\}$, reference structure $\mathbf{Z}_\mathrm{ref}$, $is\_rigid\_receptor$, $train\_plddt\_head$, loss weights $ p_\mathrm{contact}, \lambda_\mathrm{FAPE}, \lambda_\mathrm{dRMSD}, \lambda_\mathrm{viol} $ 
\FOR{$i \in \{1, \cdots, N_\mathrm{training\_iter}\}$}
    \IF{$is\_rigid\_receptor$}
        \STATE $\mathbf{x}_\mathrm{template} \gets \mathbf{x}_\mathrm{ref}$
        \ELSE
        \STATE $use\_template \sim \mathrm{Bern}(0.5)$
        \IF{$use\_template$} 
            \STATE $\mathbf{x}_\mathrm{template} \gets $ Retrieve a random structure from all AF2 predictions and PL2019-74k structures of $\{\mathbf{s}\}$
            \ELSE
            \STATE $\mathbf{x}_\mathrm{template} \gets $ None
        \ENDIF
    \ENDIF
    \IF{not ($train\_plddt\_head$ and $(i \mid 10)$)}
    \STATE Compute MHT embeddings for all input ligand molecular graphs in $\{{G}\}$ and standard amino acids \COMMENT{Algorithm~\ref{alg:MHT}}
    \STATE $N_\mathrm{P} \gets \mathrm{min}(N_\mathrm{res}, 96)$,  $N_\mathrm{L} \sim \{ \mathrm{floorRound}(\sqrt{N_\mathrm{frames}}), \cdots , 33,  32\}$
    \STATE $is\_prior\_training \sim \mathrm{Bern}(p_\mathrm{contact})$
    \IF{$is\_prior\_training$}
        \STATE $\tau \gets 1.0$, $k \sim \{0, \cdots, N_\mathrm{L} - 1\}$
        \ELSE 
        \STATE $\tau \sim \mathrm{Uniform}(0, 1)$, $k \gets N_\mathrm{L}$
    \ENDIF
    \STATE $t \gets \mathrm{DiffusionTimeSchedule}(\tau)$ \COMMENT{Equation~\eqref{eq:DiffusionTimeSchedule}}
    \STATE $\mathbf{z}_t \gets$ Perturb the structure using the forward SDE transition kernel $q_{t:0}(\cdot | \mathbf{Z}_\mathrm{ref})$ \COMMENT{Equation~\eqref{eq:forward_kernel}}
    \STATE $residue\_graph, atomic\_graph \gets$ Generate the residue-scale and atomic-scale graph based on $\mathbf{Z}_t$ \COMMENT{Section~\ref{sec:graph_construction}}
    \STATE $\mathbf{D}, \mathbf{L} \gets$ Compute ground-truth distance and contact map based on $\mathbf{Z}_\mathrm{ref}$ \COMMENT{Equation~\eqref{eq:ContactMap}, $\mathbf{D}, \mathbf{L} \in \mathbb{R}^{N_\mathrm{res}\times N_\mathrm{L} }$}
    \STATE $\tilde{\mathbf{l}} \gets  \mathbf{0}$  
    \FOR{$r \in \{1, \cdots, k\}$}
        \STATE $\mathbf{L}_r \gets \mathrm{SumIntoPatches}(L) \odot [1 - \mathrm{max}_\mathrm{column-wise}(\tilde{\mathbf{l}})]$ \COMMENT{$\mathbf{L}_r \in \mathbb{R}^{N_\mathrm{P}\times N_\mathrm{L} }$}
        \STATE $\mathbf{l}_r = \mathrm{OneHot}(i_r, j_r); \ (i_r, j_r)\sim\mathrm{Categorical}_{N_\mathrm{P}\times N_\mathrm{L}}(\mathbf{L}_r)$  \COMMENT{$\mathbf{l}_r \in \{0, 1\}^{N_\mathrm{P}\times N_\mathrm{L} }$}
        \STATE $\tilde{\mathbf{l}} \gets \tilde{\mathbf{l}} + \mathbf{l}_r$  \COMMENT{$\tilde{\mathbf{l}} \in \{0, 1\}^{N_\mathrm{P}\times N_\mathrm{L} }$}
    \ENDFOR
    \STATE $residue\_graph \gets \mathrm{CPMForward}(residue\_graph, \tilde{\mathbf{l}}, \tau)$ \COMMENT{Section~\ref{sec:CPM}}
    \STATE $\hat{P} = \mathrm{LinearNoBias}(\mathrm{MLP}(\mathrm{GetEdges_{BF}}(residue\_graph\_k)))$ \COMMENT{$\hat{P} \in \mathbb{R}^{N_\mathrm{res}\times N_\mathrm{L} \times N_\mathrm{bins}}, N_\mathrm{bins}=32$}
    \STATE $graph\_rep \gets \mathrm{Collate}(residue\_graph, atomic\_graph, cross\_scale\_graph)$ \COMMENT{Table~\ref{table:all_edges}}
    \STATE $\hat{\mathbf{Z}}_0 \gets \mathrm{ESDMForward}(\mathbf{Z}_t; graph\_rep, \tau)$ \COMMENT{Section~\ref{sec:ESDM}} \\
    \textit{\# Distogram and contact prediction losses}
    \STATE $\mathcal{L}_\mathrm{dgram}  = \mathrm{mean}_{A, J} \, \mathcal{L}_\mathrm{CE}(\hat{P}_{AJ}, \mathrm{bucketize\_onehot}(D_{AJ}, \mathbf{v}_\mathrm{bins}))$ 
    \STATE $\mathcal{L}_\mathrm{cmap} = \mathcal{L}_\mathrm{CE}(\mathrm{DistogramToContactMap}(\hat{P}), \mathbf{L})$ \\
    \textit{\# Invariant denoising structure prediction losses}
    \STATE $\mathcal{L}_\mathrm{FAPE} = \mathrm{FAPE}(\hat{\mathbf{Z}}_0, \mathbf{Z}_\mathrm{ref}) + \mathrm{FAPE}_\mathrm{scaled}(\hat{\mathbf{Z}}_0, \mathbf{Z}_\mathrm{ref})$ 
    \STATE $\mathcal{L}_\mathrm{dRMSD} = \mathrm{dRMSD}_\mathrm{global}(\hat{\mathbf{Z}}_0, \mathbf{Z}_\mathrm{ref}) + \mathrm{dRMSD}_\mathrm{site}(\hat{\mathbf{Z}}_0, \mathbf{Z}_\mathrm{ref}) + \mathrm{dRMSD}_\mathrm{pli}(\hat{\mathbf{Z}}_0, \mathbf{Z}_\mathrm{ref})  + \mathrm{dRMSD}_\mathrm{weighted}(\hat{\mathbf{x}}_0, \mathbf{x}_\mathrm{ref}; \mathbf{x}_\mathrm{template})$ 
    \STATE $\lambda(t) = \frac{\sqrt{\sigma(t)^2 + \sigma^2_\mathrm{data}}}{\sigma(t) \cdot \sigma_\mathrm{data}} ,\ \sigma(t) = \sigma \sqrt{t}$ \COMMENT{Loss weighting adapted from Ref.~\cite{karras2022elucidating}, $\sigma=\SI{12.25}{\angstrom}, \sigma_\mathrm{data}=\SI{5.0}{\angstrom}$}\\
    \textit{\# Distance-geometry and clash regularizers}
    \STATE $\mathcal{L}_\mathrm{distgeom}, \mathcal{L}_\mathrm{clash} \gets $ Compute structure violation losses \COMMENT{See Equations~\eqref{eq:distgeom_loss}-\eqref{eq:clash_loss}}
    \STATE $\mathcal{L} \gets 0.1 \cdot (\mathcal{L}_\mathrm{dgram} + \mathcal{L}_\mathrm{cmap}) + \lambda(t) \lambda_\mathrm{FAPE} \cdot \mathcal{L}_\mathrm{FAPE} + \lambda(t)  \lambda_\mathrm{dRMSD} \cdot \mathcal{L}_\mathrm{dRMSD} + \lambda(t) \lambda_\mathrm{viol} \cdot (\mathcal{L}_\mathrm{distgeom} + 0.1 \cdot \mathcal{L}_\mathrm{clash})$
    \STATE $\mathcal{L} \gets is\_prior\_training \cdot (\mathcal{L}_\mathrm{dgram} + \mathcal{L}_\mathrm{cmap}) + (1 - 0.9 \cdot is\_prior\_training) \cdot \mathcal{L}$
    \ELSE 
        \STATE \textit{\# Confidence estimation losses}
        \STATE $\mathbf{Z}_0, (pLDD\_gram, \_, \_) \gets \mathrm{NeuralPLexerInference}(\{\mathbf{s}\}, \{{G}\}, 1, N_\mathrm{steps}=10, use\_template, \mathrm{True})$ \COMMENT{Alg.~\ref{alg:main}}
        \STATE $LDD\_gram \gets$ Compute reference distance deviation histograms between $\mathbf{Z}_0$ and $\mathbf{Z}_\mathrm{ref}$ \COMMENT{See Section~\ref{sec:si_plddt} text}
        \STATE $\mathcal{L} \gets \mathrm{mean}_{A\in \{\mathbf{s}\}} \mathcal{L}_\mathrm{CE}(pLDD\_gram_A, LDD\_gram_A) + \mathrm{mean}_{J\in \{{G}\}} \mathcal{L}_\mathrm{CE}(pLDD\_gram_J, LDD\_gram_J)$ 
    \ENDIF
    \STATE Computing gradients w.r.t. $\mathcal{L}$ and update model parameters
\ENDFOR
\end{algorithmic}
\end{algorithm}

$\mathrm{FAPE}$ denotes the frame-aligned point error introduced in Ref.~\cite{jumper2021highly} for which we used all ${N}_\mathrm{res}+{N}_\mathrm{frame}$ backbone and ligand frames for relative point coordinate alignment; $\mathrm{FAPE}_\mathrm{scaled}$ is a modified variant of FAPE to encourage learning the overall molecular chirality, where we scaled all aligned point coordinates by its vector norm and removed the clamping term.

$\mathrm{dRMSD}_\mathcal{X}$ denotes the distance-based RMSD introduced in Ref.~\cite{alquraishi2019end} computed for all atoms in a selected set $\mathcal{X}$; we denote $\mathcal{X}=\mathrm{global}$ for all atoms in the protein-ligand complex,  $\mathcal{X}=\mathrm{site}$ for all atoms within \SI{8.0}{\angstrom} of the binding ligands, $\mathcal{X}=\mathrm{pli}$ for all pairs between protein residue centroids and ligand atoms, and $\mathcal{X}=\mathrm{weighted}$ for all pairs for which the residue-residue distance deviation between the reference structure $\mathbf{x}_\mathrm{ref}$ and the template structure $\mathbf{x}_\mathrm{template}$ is greater than \SI{2.0}{\angstrom}.

The distance-geometry loss $\mathcal{L}_\mathrm{distgeom}$ is defined as
\begin{equation}
    \label{eq:distgeom_loss}
    \mathcal{L}_\mathrm{distgeom}( \hat{\mathbf{Z}}, \mathbf{Z}) = \mathrm{mean}_A \sum_{i, j \in \{A\}} | \lVert \mathbf{Z}_i - \mathbf{Z}_j \rVert - \lVert \hat{\mathbf{Z}}_i - \hat{\mathbf{Z}}_j \rVert | \cdot \mathbbm{1}_{ \lVert \mathbf{Z}_i - \mathbf{Z}_j \rVert < \SI{2.8}{\angstrom}}
\end{equation}
where $\mathbbm{1}$ denotes a 0-1 indicator function, all coordinates are in the Angstrom unit, index $A$ runs over all residues and ligand molecules in the structure, and $i, j \in \{A\}$ indicates all atom pairs in the residue or ligand $A$.

The steric clash loss $\mathcal{L}_\mathrm{clash}$ is defined as
\begin{equation}
    \label{eq:clash_loss}
    \mathcal{L}_\mathrm{clash}( \hat{\mathbf{Z}}, \mathbf{Z}) = \sum_{i,j} \mathrm{max}( \sigma_{ij} - | \lVert \hat{\mathbf{Z}}_i - \hat{\mathbf{Z}}_j \rVert |, 0) \cdot  \mathbbm{1}_{ \lVert \mathbf{Z}_i - \mathbf{Z}_j \rVert  > 1.2 \sigma_{ij}}.
\end{equation}
where $\sigma_{ij} = \frac{1}{2}(\sigma_i + \sigma_j)$ is the average VdW radius of atom $i$ and atom $j$.

The NeuralPLexer main network (including the MHT (51.6M), CPM (8.3M), ESDM (5.0M), and projection layers) has $65.8 \times 10^6$ trainable parameters in total, while the pLDDT head has $5.0 \times 10^6$ parameters in addition.

\begin{table}
    \centering
    \caption{Model training details. All training runs were performed on 6 NVIDIA-Tesla-V100-SXM2-32GB GPUs with a total batch size of 6, data parallelism, and automatic mixed precision (AMP) training. CA: Cosine Annealing. EMA: Exponential Moving Average.}
    \label{table:training}
    \small
    \begin{tabular}{lcccc}
        \toprule
        Task type & \multicolumn{2}{c}{End-to-end structure prediction} & Rigid-backbone docking & Binding site recovery \\
        \cmidrule(lr){2-3} \cmidrule(lr){4-4} \cmidrule(lr){5-5} 
        Model ID & A.0 & A.1 & B & C \\
        \midrule
        Initial parameters & Random & A.0 & A.0 & A.1 \\
        Training dataset & \begin{tabular}{@{}c@{}}PL2019-74k \\ (DNA/RNA unremoved)\end{tabular}  & PL2019-74k & PDBBind2020(<2019)  & PDBBind2020(<2019) \\
        Data sampling scheme & see text & see text & N/A & N/A \\
        $is\_rigid\_receptor$ & no & no & yes & yes (cropped binding site) \\
        $train\_plddt\_head$ & no & yes & yes & no \\
        Freeze MHT parameters & no & yes & yes & yes \\
        Initial learning rate & $3\cdot 10^{-4}$ & $1 \cdot 10^{-4}$ &  $2 \cdot 10^{-4}$ &  $2 \cdot 10^{-4}$\\
        Learning rate schedule & CA, 4 restarts & CA & CA, 6 restarts & CA, 4 restarts \\
        Dropout rate & 0.1 & 0.01 & 0.01 & 0.01 \\
        Template masking rate & 0.2 & 0.2 & 0.2 & 0.2 \\
        Use EMA & no & yes & yes & no \\
        $p_\mathrm{contact}$ & 0.25 & 0.1 & 0.1 & 0.1 \\
        $\lambda_\mathrm{FAPE}$ & 0.5 & 0.2 & 0.2 & 0.2 \\
        $\lambda_\mathrm{dRMSD}$ & 0.5 & 1.0 & 1.0 & 1.0 \\
        $\lambda_\mathrm{viol}$ & $10^{-3}$ & linear increase, 0 to 0.1 & linear increase, 0 to 0.1 & 0.1 \\
        Num. epochs & 75 + 250 + 250 + 224 & 332 & 36 + 4 + 8 + 16 + 32 + 60  & 7 + 14 + 28 + 56 \\
        Training time & 10 days 20 hours & 5 days 8 hours & 7 days 16 hours & 3 days 12 hours\\
        \bottomrule
    \end{tabular}
\end{table}

In Table~\ref{table:training} we summarize the training procedure and hyperparameters for all models reported in this study. During the training of model A.0, we randomly subsampled 5\% of the entire PL2019-74k dataset at each epoch using a relative weight of the inverse-square-root of the UniRef50 cluster index occurrence frequency of the training protein chain. 
During the training of model A.1, we subsampled 5\% of the PL2019-74k dataset  at each epoch with an additional relative sampling weight of (a) $(1 + 0.5 * \exp(1.0 * \mathrm{BackboneRMSD}(\mathbf{x}_\mathrm{ref}, \mathbf{x}_\mathrm{template})))$ for AF2 templates and (b) $(1 + 0.3 * \exp(1.1 * \mathrm{BackboneRMSD}(\mathbf{x}_\mathrm{ref}, \mathbf{x}_\mathrm{template})))$ for experimental templates, multiplied to the sampling weights used for training model A.0. All PDBBind fine-tuning runs (models B and C) were executed with standard data shuffling and no subsampling.

\section{Computational details}

For results reported in blind protein-ligand docking and binding site recovery tasks, we used a diffusion model sampling setting of $N_\mathrm{step}=25$ integrator steps; for all end-to-end structure prediction results, we used $N_\mathrm{step}=100$ integrator steps.

The AlphaFold2 structures used for binding site structure recovery (Figure~\ref{fig:fig3}) and for templates in end-to-end structure prediction (Figure~\ref{fig:fig4}-\ref{fig:fig5}) are predicted using ColabFold~\cite{mirdita2022colabfold} using %
default recycling and AMBER relaxation settings, and without templates in order to best reflect the prediction fidelity of AlphaFold2 on new targets. %
The input sequences for all protein chains are obtained from \url{https://www.ebi.ac.uk/pdbe/api/pdb/entry/molecules/} to avoid issues related to unresolved residues and to represent a realistic testing scenario where the protein backbone models are obtained from the full sequence.

In Figure~\ref{fig:fig3}b, EquiBind~\cite{stark2022equibind} is launched with the default configuration file, and for each protein-ligand pair 16 ligand conformations are generated using different random RDKit~\cite{landrum2013rdkit} input conformers. %

RosettaLigand~\cite{davis2009rosettaligand} runs are launched with a configuration modified from the standard protocol. We set the receptor Calpha constraint parameter to 100.0 to enable a fully flexible receptor; the ligand coordinates are initialized using the aligned-ground-truth conformation as obtained by TM-Align~\cite{zhang2005tm}, with randomized torsion angles using the BCL~\cite{brown2022introduction} library as described in the standard protocol. We set the docking box width to $\SI{4.0}{\angstrom}$ and remove the ligand center perturbation step to ensure the ligand search space during the low-resolution docking stage is constrained to the binding site location.  %
We set the number of docking cycles of the high-resolution docking stage to 64 to converge the receptor structure during backbone relaxation; for each protein-ligand pair in the test set, generating 32 ligand poses took on average 150 minutes on a single Intel(R) Xeon(R) CPU E5-2698 v4 @ 2.20GHz CPU.

All protein structure alignments and TM-score calculations are performed using TMAlign~\cite{zhang2005tm}. All reported TM-scores are normalized by the chain length of the reference PDB structure. The per-residue all-atom lDDT score is computed using OpenStructure~\cite{biasini2013openstructure}  with an inclusion radius of $\SI{10.0}{\angstrom}$; the lDDT-BS score is then computed by averaging the per-residue scores for binding site residues within $\SI{4.0}{\angstrom}$ of the ligand. The symmetry-corrected heavy-atom RMSD for ligand structure comparison is computed using the $\mathrm{obrms}$ function in OpenBabel~\cite{o2011open}. A standard 6-12 Lennard-Jones energy functional form is used for computing the clash rate statistics; the L-J energy and VdW radius parameters are obtained from the UFF parameter file retrieved from \url{https://github.com/kbsezginel/lammps-data-file/blob/master/uff-parameters.csv}.

\section{Supplementary results}

In Tables~\ref{table:pred_pocketminer}-\ref{table:pred_recent} we summarize the performance statistics for end-to-end structure prediction on targets reported in this study. For each target, we generate six AF2 structure models with the following setup:
\begin{itemize}
    \item For model 0, we use an MSA cluster size = 256, extra MSA size = 512, 3 recycling cycles, and the AF2 model version "AlphaFold-ptm:3". This set of results is also used to generate the visualizations in Figure~\ref{fig:fig4}a-c.
    \item For models 1-5, we use an MSA cluster size = 512, extra MSA size = 1024, 3 recycling cycles, and AF2 model versions "AlphaFold-ptm:1", "AlphaFold-ptm:2", "AlphaFold-ptm:3", "AlphaFold-ptm:4", "AlphaFold-ptm:5", respectively.
\end{itemize}

These AF2 structure models are used as the template inputs for NeuralPLexer to generate six independent sets of predictions; for each apo protein or protein-ligand pair, we sample 8 NeuralPLexer structures using independent random seeds. On both datasets, NeuralPLexer achieves the highest average prediction accuracy (as shown by the highest TM-score and lowest backbone RMSD).

We observe that NeuralPLexer achieves a further improvement against AF2 when averaged against the sampled structures of the highest TM-score for each target (Tables~\ref{table:pred_pocketminer}-\ref{table:pred_recent}, "per-target top-1"), suggesting that the multi-structure generative model formulation naturally enables better coverage of the experimental structure compared to heuristic approaches that are based on randomizing AF2.

\begin{table}[htp]
\caption{Model predictions on the PocketMiner dataset (33 holo structures + 29 apo structures).}
\label{table:pred_pocketminer}
\footnotesize
\begin{tabular}{@{}lllcccccc@{}}
\toprule
Method &
  Sampler &
   &
  \begin{tabular}[c]{@{}l@{}}TM-score\\ (all predictions)\end{tabular} &
  \begin{tabular}[c]{@{}l@{}}Backbone RMSD\\ (all predictions)\end{tabular} &
  \begin{tabular}[c]{@{}l@{}}TM-score\\ (per-target top-1)\end{tabular} &
  \begin{tabular}[c]{@{}l@{}}Backbone RMSD\\ (per-target top-1)\end{tabular} &
  \begin{tabular}[c]{@{}l@{}}Ligand RMSD\\ (all predictions)\end{tabular} &
  \begin{tabular}[c]{@{}l@{}}Ligand RMSD\\ (per-target top-1)\end{tabular} \\
  \midrule
\multirow{5}{*}{AlphaFold2}                                                          & \multirow{5}{*}{-}       & mean & 0.929 & 1.548 & 0.943 & 1.323 & -     & -      \\
                                                                                     &                          & std  & 0.095 & 0.813 & 0.090 & 0.721 & -     & -      \\
                                                                                     &                          & 25\% & 0.931 & 0.897 & 0.948 & 0.755 & -     & -      \\
                                                                                     &                          & 50\% & 0.962 & 1.350 & 0.969 & 1.095 & -     & -      \\
                                                                                     &                          & 75\% & 0.978 & 2.032 & 0.986 & 1.710 & -     & -      \\ \midrule
\multirow{10}{*}{\begin{tabular}[c]{@{}l@{}}NeuralPLexer\\ (no ligand)\end{tabular}} & \multirow{5}{*}{DDIM~\cite{song2020denoising}}    & mean & 0.895 & 2.082 & 0.938 & 1.501 & -     & -      \\
                                                                                     &                          & std  & 0.092 & 0.778 & 0.079 & 0.567 & -     & -      \\
                                                                                     &                          & 25\% & 0.877 & 1.480 & 0.934 & 1.062 & -     & -      \\
                                                                                     &                          & 50\% & 0.920 & 1.950 & 0.959 & 1.470 & -     & -      \\
                                                                                     &                          & 75\% & 0.951 & 2.620 & 0.973 & 1.812 & -     & -      \\ \cmidrule{2-9}
                                                                                     & \multirow{5}{*}{LSA-SDE} & mean & 0.925 & 1.623 & 0.946 & 1.305 & -     & -      \\
                                                                                     &                          & std  & 0.091 & 0.784 & 0.087 & 0.647 & -     & -      \\
                                                                                     &                          & 25\% & 0.910 & 0.970 & 0.952 & 0.788 & -     & -      \\
                                                                                     &                          & 50\% & 0.958 & 1.470 & 0.970 & 1.195 & -     & -      \\
                                                                                     &                          & 75\% & 0.973 & 2.172 & 0.982 & 1.625 & -     & -      \\ \midrule
\multirow{10}{*}{\begin{tabular}[c]{@{}l@{}}NeuralPLexer\\ (ours)\end{tabular}}      & \multirow{5}{*}{DDIM~\cite{song2020denoising}}    & mean & 0.909 & 1.893 & 0.943 & 1.387 & 9.115 & 10.00  \\
                                                                                     &                          & std  & 0.089 & 0.684 & 0.084 & 0.559 & 9.466 & 11.11 \\
                                                                                     &                          & 25\% & 0.898 & 1.370 & 0.943 & 1.008 & 3.101 & 3.394  \\
                                                                                     &                          & 50\% & 0.933 & 1.780 & 0.967 & 1.260 & 5.398 & 5.390  \\
                                                                                     &                          & 75\% & 0.958 & 2.310 & 0.978 & 1.680 & 9.808 & 10.30 \\ \cmidrule{2-9}
                                                                                     & \multirow{5}{*}{LSA-SDE} & mean & \textbf{0.934} & \textbf{1.486} & \textbf{0.950} & \textbf{1.236} & 8.985 & 9.812  \\
                                                                                     &                          & std  & 0.091 & 0.750 & 0.087 & 0.656 & 9.943 & 10.75 \\
                                                                                     &                          & 25\% & 0.938 & 0.900 & 0.955 & 0.720 & 2.674 & 2.583  \\
                                                                                     &                          & 50\% & 0.966 & 1.290 & 0.975 & 1.095 & 5.010 & 5.246  \\
                                                                                     &                          & 75\% & 0.978 & 1.870 & 0.987 & 1.613 & 9.388 & 10.54 \\
                                                                                     \bottomrule
\end{tabular}
\end{table}

\begin{table}[htp]
\caption{Model predictions on 118 recent targets with ligand-induced conformational changes.}
\label{table:pred_recent}
\footnotesize
\begin{tabular}{@{}lllcccccc@{}}
\toprule
Method &
  Sampler &
   &
  \begin{tabular}[c]{@{}l@{}}TM-score\\ (all predictions)\end{tabular} &
  \begin{tabular}[c]{@{}l@{}}Backbone RMSD\\ (all predictions)\end{tabular} &
  \begin{tabular}[c]{@{}l@{}}TM-score\\ (per-target top-1)\end{tabular} &
  \begin{tabular}[c]{@{}l@{}}Backbone RMSD\\ (per-target top-1)\end{tabular} &
  \begin{tabular}[c]{@{}l@{}}Ligand RMSD\\ (all predictions)\end{tabular} &
  \begin{tabular}[c]{@{}l@{}}Ligand RMSD\\ (per-target top-1)\end{tabular} \\ \midrule
\multirow{5}{*}{AlphaFold2}                                                          & \multirow{5}{*}{-}       & mean & \textbf{0.891} & \textbf{2.049} & 0.911 & 1.881 & -      & -      \\
                                                                                     &                          & std  & 0.111 & 1.007 & 0.099 & 1.057 & -      & -      \\
                                                                                     &                          & 25\% & 0.852 & 1.368 & 0.865 & 1.103 & -      & -      \\
                                                                                     &                          & 50\% & 0.930 & 1.910 & 0.951 & 1.615 & -      & -      \\
                                                                                     &                          & 75\% & 0.969 & 2.592 & 0.975 & 2.322 & -      & -      \\ \midrule
\multirow{10}{*}{\begin{tabular}[c]{@{}l@{}}NeuralPLexer\\ (no ligand)\end{tabular}} & \multirow{5}{*}{DDIM~\cite{song2020denoising}}    & mean & 0.844 & 2.695 & 0.908 & 1.995 & -      & -      \\
                                                                                     &                          & std  & 0.112 & 0.924 & 0.092 & 0.878 & -      & -      \\
                                                                                     &                          & 25\% & 0.802 & 2.010 & 0.898 & 1.390 & -      & -      \\
                                                                                     &                          & 50\% & 0.877 & 2.570 & 0.942 & 1.755 & -      & -      \\
                                                                                     &                          & 75\% & 0.920 & 3.270 & 0.961 & 2.370 & -      & -      \\ \cmidrule{2-9}
                                                                                     & \multirow{5}{*}{LSA-SDE} & mean & 0.877 & 2.261 & 0.916 & 1.847 & -      & -      \\
                                                                                     &                          & std  & 0.110 & 1.018 & 0.093 & 1.005 & -      & -      \\
                                                                                     &                          & 25\% & 0.833 & 1.580 & 0.891 & 1.218 & -      & -      \\
                                                                                     &                          & 50\% & 0.916 & 2.070 & 0.949 & 1.585 & -      & -      \\
                                                                                     &                          & 75\% & 0.950 & 2.790 & 0.971 & 2.173 & -      & -      \\ \midrule
\multirow{10}{*}{\begin{tabular}[c]{@{}l@{}}NeuralPLexer\\ (ours)\end{tabular}}      & \multirow{5}{*}{DDIM~\cite{song2020denoising}}    & mean & 0.867 & 2.418 & 0.921 & 1.818 & 14.22 & 13.79 \\
                                                                                     &                          & std  & 0.110 & 0.908 & 0.089 & 0.948 & 13.01 & 12.50 \\
                                                                                     &                          & 25\% & 0.839 & 1.730 & 0.918 & 1.162 & 3.567  & 3.156  \\
                                                                                     &                          & 50\% & 0.903 & 2.280 & 0.953 & 1.555 & 7.410  & 7.878  \\
                                                                                     &                          & 75\% & 0.941 & 2.940 & 0.971 & 2.218 & 23.55 & 24.27 \\ \cmidrule{2-9}
                                                                                     & \multirow{5}{*}{LSA-SDE} & mean & \textbf{0.893} & \textbf{2.026} & \textbf{0.926} & \textbf{1.676} & 14.03 & 12.72 \\
                                                                                     &                          & std  & 0.110 & 0.982 & 0.091 & 0.961 & 12.94 & 12.23 \\
                                                                                     &                          & 25\% & 0.866 & 1.360 & 0.916 & 1.055 & 3.353  & 3.293  \\
                                                                                     &                          & 50\% & 0.929 & 1.870 & 0.958 & 1.425 & 8.146  & 5.907  \\
                                                                                     &                          & 75\% & 0.965 & 2.570 & 0.978 & 2.118 & 22.97 & 21.60 \\ \bottomrule
\end{tabular}
\end{table}

\clearpage

\printbibliography
\end{refsection}

\end{document}